\documentclass[trackchanges]{aastex701}
\usepackage[T1]{fontenc}
\usepackage{newtxtext,newtxmath}
\usepackage{graphicx}	
\usepackage{amsmath}	
\usepackage{subcaption}
\newcommand{\teff}{$T_{\rm eff}$}
\newcommand{\logg}{$\log g$}
\newcommand{\vsini}{$v \sin i$}
\newcommand{\kms}{km\,s$^{-1}$}
\newcommand{\ds}{$\delta$ Scuti}

\begin{document}

\title{Pulsating and Non-pulsating Components of Detached Eclipsing Binaries in the \ds\, instability strip}

\author[orcid=0000-0002-9036-7476,sname='Kahraman Ali\c{c}avu\c{s}']{Filiz Kahraman Ali\c{c}avu\c{s}}
\affiliation{\c{C}anakkale Onsekiz Mart University, Faculty of Sciences, Physics Department, 17100, \c{C}anakkale, T\"{u}rkiye}
\affiliation{\c{C}anakkale Onsekiz Mart University, Astrophysics Research Center and Ulup{\i}nar Observatory, TR-17100, \c{C}anakkale, T\"{u}rkiye}
\email[show]{filizkahraman01@gmail.com}  

\author[orcid=0000-0002-1972-8400, sname='Ali\c{c}avu\c{s}']{Fahri Ali\c{c}avu\c{s}} 
\affiliation{\c{C}anakkale Onsekiz Mart University, Faculty of Sciences, Physics Department, 17100, \c{C}anakkale, T\"{u}rkiye}
\affiliation{\c{C}anakkale Onsekiz Mart University, Astrophysics Research Center and Ulup{\i}nar Observatory, TR-17100, \c{C}anakkale, T\"{u}rkiye}
\email{fahrilcvs@gmail.com}

\author[orcid=0000-0002-9424-2339, sname='\c{C}elik Orhan']{Zeynep \c{C}elik Orhan} 
\affiliation{Department of Astronomy and Space Sciences, Faculty of Science, Ege University, 35100, \.{I}zmir, T\"{u}rkiye}
\email{zcelik87@gmail.com }

\author[0000-0003-3278-5140,sname='\c{C}elik']{Eda \c{C}elik}
\affiliation{\c{C}anakkale Onsekiz Mart University, School of Graduate Studies, Department of Physics, TR-17100, \c{C}anakkale, T\"{u}rkiye}
\email{celikeda7575@gmail.com }

\author[0000-0003-4337-8612,sname='Catanzaro']{Giovanni Catanzaro}
\affiliation{INAF–Osservatorio Astrofisico di Catania, Via S. Sofia 78, I-95123 Catania, Italy}
\affiliation{Dipartimento di Fisica e Astronomia, Sezione Astrofisica, Università di Catania, Via S. Sofia 78, I-95123 Catania, Italy}
\email{giovanni.catanzaro@inaf.it }

\author[0000-0002-4453-1597,sname='Giarrusso']{Marina Giarrusso}
\affiliation{INAF–Osservatorio Astrofisico di Catania, Via S. Sofia 78, I-95123 Catania, Italy}
\email{marina.giarrusso@inaf.it}


\begin{abstract}

Pulsating detached eclipsing binary systems are crucial for studying the internal structure of oscillating stars. These systems are advantageous because binary effects on pulsations are minimal, allowing for more accurate determinations of fundamental stellar parameters such as mass and radius. They serve as unique laboratories for detailed investigations of pulsating stars. In this study, we focused on four detached eclipsing binaries exhibiting $\delta$ Scuti-type oscillations: HD\,117476, 205\,Dra, HY\,Vir, and V1031\,Ori. Our preliminary investigation showed that all binary components of these targets lie within the $\delta$ Scuti instability strip. Therefore, we aimed to determine which components are pulsating and which are not, and to explore the differences between them. To achieve this, we analyzed TESS photometric data and high-resolution spectra of the targets. Radial velocity variations were measured, and atmospheric parameters for each component were derived using spectral disentangling or synthetic composite spectra. We also modeled the binary light and radial velocity curves to determine the fundamental physical parameters of the components. Furthermore, we examined pulsation properties using three different approaches to identify the pulsating components. The evolutionary status of the targets was also assessed. Our analysis revealed that, in each system, only one component exhibits $\delta$ Scuti-type pulsations, while the others are non-pulsating. Interestingly, we found that the key difference between pulsating and non-pulsating components within the same binary is metallicity: the metal-rich components were found to be non-pulsators, supporting theoretical studies on the effect of metallicity on $\delta$ Scuti-type pulsations.

\end{abstract}

\keywords{stars: binaries: eclipsing -- -- stars: variables: delta Scuti -- stars: fundamental parameters}


\section{Introduction}

Space-based telescopes have facilitated the discovery of numerous unique variable stars, including eclipsing binaries and pulsating variables \citep{2019MNRAS.490.4040A, 2021Univ....7..369S, 2022afas.confE...1K,2022MNRAS.510.1413K, 2023MNRAS.524..619K}. Moreover, aside from these, many new candidates of variable stars exhibiting different variability simultaneously have been unveiled, such as hybrid pulsators and eclipsing binaries with pulsating components \citep{2019MNRAS.490.4040A, 2022ApJS..263...34C}. These discoveries have furnished us with high-quality data, enabling a detailed understanding of stars' structure and evolution.

Among these variables, eclipsing binary systems with oscillating stars stand out as particularly remarkable. Eclipsing binary systems, especially the detached binaries, offer a means to accurately estimate fundamental stellar parameters such as mass ($M$) and radius ($R$) with high precision (<1\%, \citeauthor{2024A&A...691A.170H} \citeyear{2024A&A...691A.170H}, \citeauthor{2022A&A...666A.128G} \citeyear{2022A&A...666A.128G}, \citeauthor{2013A&A...557A.119S} \citeyear{2013A&A...557A.119S}, \citeauthor{2010A&ARv..18...67T} \citeyear{2010A&ARv..18...67T}). 
However, all eclipsing binaries are not suited for investigating stellar interiors; instead, the oscillations of pulsating stars provide insight into the stellar interior. Therefore, eclipsing binaries, especially those with at least one pulsating component, play a crucial role in deepening our understanding of stellar systems. Given that the components of eclipsing binaries can vary widely in spectral type, ranging from approximately B to G \citep{2015ASPC..496..164S}, and may also display different evolutionary statuses, it is conceivable to encounter a scenario where one of these binary components falls within the instability strip of a pulsating star and exhibits oscillations. The $\beta$\,Cephei, $\delta$\,Scuti and $\gamma$\,Doradus pulsators are some of these oscillating systems found as a member of eclipsing binaries \citep{2024A&A...691A.101P, 2024MNRAS.527.4076K, 2024ApJS..272...25E, 2022afas.confE...1K, 2022MNRAS.513.3191S, 2022MNRAS.515.2755S}. Among these pulsators, $\delta$\,Scuti variables are more frequently detected as binary components, primarily due to their relatively short pulsation periods compared to other types of pulsators. \ds\, stars are systems of A to F types that display oscillations typically ranging from 18 minutes to 8 hours, primarily in pressure modes \citep{2013AJ....145..132C, 2019MNRAS.490.4040A}. In a recent catalog study of $\delta$\,Scuti variables in eclipsing binaries, the number of these systems was estimated to be around 90 \citep{2017MNRAS.470..915K}. However, recent space-based discoveries have significantly augmented the known number of these systems, suggesting that there could be approximately 400 $\delta$\,Scuti stars in eclipsing binaries in total \citep{2022ApJS..263...34C, 2022MNRAS.510.1413K, 2023MNRAS.524..619K}. 

Recent studies have revealed that the boundaries of the $\delta$\,Scuti instability strip, and the pulsation mechanism of these systems have not been fully understood \citep{2014ApJ...796..118A, 2019MNRAS.485.2380M}. Theoretically, it was expected that $\delta$\,Scuti-type oscillations would be confined within their instability domain, with no pulsations detected beyond the borders of this instability strip. However, some $\delta$\,Scuti variables have been observed outside of their instability strip \citep{2011A&A...534A.125U, 2019MNRAS.490.4040A}, and, even more surprisingly, some non-variable stars have been found within the $\delta$\,Scuti instability strip \citep{2015AstRv..11....1G, 2019MNRAS.485.2380M}. In addition, one important factor appears to be metallic A (Am) and F (Fm) stars lie within or near the \ds\, instability strip, their peculiar atmospheric abundances and slower rotation rates were traditionally thought to suppress pulsations \citep{1976ApJS...32..651K, 2000A&A...360..603T}. However, observational evidence has shown that many Am/Fm stars do pulsate \citep{2022afas.confE...1K}. For example, \cite{2017MNRAS.465.2662S} demonstrated that the incidence of \ds\,-type pulsations in Am stars decreases as chemical peculiarity (metallicism) increases, suggesting a connection between metallicity and pulsation driving. More recently, \cite{2024A&A...690A.104D} analyzed a large sample of Am and Fm stars using TESS and Gaia data and found that nearly 50\% of them show \ds\,-type oscillations. These results indicate that pulsation behavior in the \ds\, domain is influenced not only by temperature and gravity, but also by detailed chemical properties providing an important context for interpreting the presence of both pulsating and non-pulsating stars within the instability strip.
Hence, more detailed studies of these systems are necessary to gain a deeper understanding of their characteristics. For this investigation, eclipsing binaries with $\delta$\,Scuti pulsators are particularly suitable, especially those situated in detached eclipsing binaries. In detached eclipsing binary systems, the influence of binarity on pulsations, such as mass transfer, is negligible, and it is possible to determine the fundamental stellar parameters of the pulsating component with high accuracy. 

In this study, we focus on detached eclipsing binaries that exhibit $\delta$\,Scuti-type oscillations. Our preliminary investigation suggests that all components of these binaries lie within the \ds\, instability strip. We aim to determine whether the components of binary systems are pulsating, or if only one is. By doing this, we will investigate the pulsation structures of binary components that formed from the same interstellar medium and lie within the same instability strip. HD\,117476, 205\,Dra, HY\,Vir and V1031\,Ori are selected as targets for our study. The pulsational behaviors of the target systems were investigated with the space-based Transiting Exoplanet Survey Satellite (TESS) data \citep{2022RAA....22h5003K, 2023MNRAS.524..619K}. HY\,Vir and V1031\,Ori were examined with spectroscopic and ground-based photometric data by \cite{2011AJ....142..185S, 2008NewA...13..304M} and \cite{1990A&A...228..365A}, respectively. While other targets have not been subjected to detailed spectroscopic and photometric investigations in the literature, conducting updated spectroscopic and space-based photometric analysis of the selected systems will help us better understand the nature of binary components within the \ds\, instability strip. Hence, we organize the paper as follows: In Sect.\,2, information about the observational data is presented. In Sect.\,3 and 4, details about the spectroscopic and photometric analyses are provided. The evolutionary modeling of all systems is introduced in Sect.\,5. Finally, in Sect.\,6, the discussion and conclusions are presented.

\section{Observational data}

For the detailed investigation of the selected detached eclipsing binary systems, both photometric and spectroscopic data were gathered. 

The selected systems were observed by the Transiting Exoplanet Survey Satellite \citep[TESS, ][]{2014SPIE.9143E..20R}. TESS has been observed in nearly the entire sky using a wide-band filter centered around the I wavelength and operated by dividing the sky into sectors. Each sector undergoes observation for approximately 27 days. While TESS's primary goal is the detection of new exoplanets, it has also provided high-quality data for many variable stars. The TESS data are accessible through the Barbara A. Mikulski Archive for Telescopes (MAST) portal\footnote{https://mast.stsci.edu}, where data captured with various exposure times (e.g., 120\,s, 600\,s, 1800\,s) and in different fluxes, are available. Both the simple aperture photometry (SAP) and pre-search data conditioning SAP (PDCSAP) fluxes of TESS data can be retrieved from this portal and we preferred to use SAP fluxes in our analysis. The target systems have been observed by TESS across multiple sectors with different exposure times. However, for our study, we opt to utilize the 120\,s data, as its Nyquist frequency is more conducive for detecting short-period pulsations, such as those exhibited by $\delta$\,Scuti stars. Detailed information about the target systems and their corresponding TESS data \citep{2025Tess} can be found in Table\,1.

\begin{table*}
    \centering 
      \caption{Information about the target detached eclipsing binary systems. SNR is the abbreviation of signal-to-noise ratio. The orbital period (P$_{orb}$) value of the systems were taken from \cite{2004AcA....54..207K, 2022RAA....22h5003K}. }\label{table1}
    \begin{tabular}{lllllllll}
    \hline
Star  & RA   & DEC  & V     & P$_{orb}$ &TESS    & Spectrograph & Number of & Average\\
Name  & (deg)& (deg)& (mag) & (day)     &sector &              & spectra   & SNR\\     
        \hline
HD\,117476 & 202.5 & 34.5  & 7.72 & 1.313542 (7) & 23, 49, 50     & CAOS  & 10 & 72\\
205\,Dra   & 281.9 & 49.4  & 7.18 & 4.243430 (7)& 14, 15, 26, 40  & CAOS & 8 & 70\\
HY\,Vir    & 197.1 & -2.7  & 7.86 & 2.732334 (1)& 46              & HARPS & 11 & 70\\
V1031\,Ori & 86.9  & -10.5 & 6.09 & 3.405561 (1)& 6, 33           & FEROS & 3 & 120\\
        \hline
    \end{tabular}
\end{table*}

Following the discovery of pulsational structures in our target systems by our personal research and in studies by \cite{2022RAA....22h5003K} and \cite{2023MNRAS.524..619K}, we searched public spectral databases and found that HY\,Vir and V1031\,Ori have available spectral data. Specifically, we found that HY\,Vir has 11 spectra obtained using the High Accuracy Radial Velocity Planet Searcher (HARPS, R=80,000), an échelle spectrograph attached to the 3.6-m telescope at the European Southern Observatory (ESO) in La Silla, Chile \citep{2003Msngr.114...20M}. These spectra were recorded between 2009 and 2015. Three spectra of V1031\,Ori were observed by the Fibre-fed Extended Range Optical Spectrograph (FEROS, R=48,000) that is an \'{e}chelle spectrograph installed on the 2.2-meter telescope at the ESO in La Silla, Chile \citep{2012MNRAS.420.2727E}. Observations of V1031\,Ori were conducted in September 2014. The public spectra of both systems were retrieved from the ESO Science Archive Facility\footnote{http://archive.eso.org/cms.html}. Additionally, we conducted spectroscopic observations on HD\,117476 and 205\,Dra using the Catania Astrophysical Observatory Spectropolarimeter (CAOS). CAOS is a high-resolution, fiber-fed, cross-dispersed \'{e}chelle spectrograph installed on the 91-cm telescope at the Catania Astrophysical Observatory, located on Mount Etna, Italy \citep{2016AJ....151..116L}. It provides spectra with a resolving power of 38,000 in the wavelength range of approximately 4100\,$-$\,6700 {\AA}.  Details regarding the spectroscopic observations are provided in Table\,1. 

Both photometric and spectroscopic data underwent preprocessing before analysis. The TESS photometric data, captured with a 120-second exposure length, were processed using SAP flux and subsequently converted into magnitudes using the method described by \cite{2022RAA....22h5003K}. The spectroscopic data underwent reduction through dedicated pipelines, following standard procedures including bias subtraction, flat correction, and wavelength calibration. These processed spectra were then normalized using the SUPPNet normalization program \citep{2022ASPC..532..251R} to prepare them for further spectroscopic analysis. 

\section{Spectroscopic analysis}

The high-quality spectroscopic data of targets were suitable for radial velocity estimation and determination of atmospheric parameters. Our spectroscopic analyses aimed to derive fundamental orbital and atmospheric parameters for each system.

\subsection{Radial velocity analysis}

Fundamental astrophysical orbital parameters were derived by estimating the radial velocity (RV) shift of eclipsing binary systems. Initially, the RV values were measured using the spectra of the systems. To estimate the RV shift, we employed the FXCOR task from the IRAF\footnote{http://iraf.noao.edu/} program package \citep{1986SPIE..627..733T}. FXCOR applies a cross-correlation technique, comparing the observed spectrum with a template spectrum to determine the RV shift. Hence, at least one template spectrum is required, which could either be the spectrum of a radial velocity standard star or a synthetic spectrum.

For our analysis, we utilized synthetic spectra as templates. These synthetic spectra were generated using the ATLAS9 model atmospheres \citep{1993KurCD..13.....K} with the SYNTHE code \citep{1981SAOSR.391.....K}. To generate the theoretical spectra, we considered the effective temperature (\teff) values of systems given in the literature or the TESS Input Catalog \citep[TIC;][]{2019AJ....158..138S}\footnote{For HD 117476, 205\,Dra, the TIC \teff\, values of 7737 and 6814\,K were considered. For HY\,Vir and V1031\,Ori, 6855 and 8797\,K were taken into account respectively \citep{2011AJ....142..185S,
 2021PASJ...73..809L}.}. These theoretical spectra served as templates during the RV analysis. The measured RV values are presented in Table\,\ref{tab:A1}. All targets were found to be double systems expect for V1031\,Ori which exhibit the variation of a third system in the spectra used in our analysis.

\begin{figure*}
\centering
\begin{subfigure}{1\textwidth}
  \centering
  \includegraphics[width=.45\linewidth]{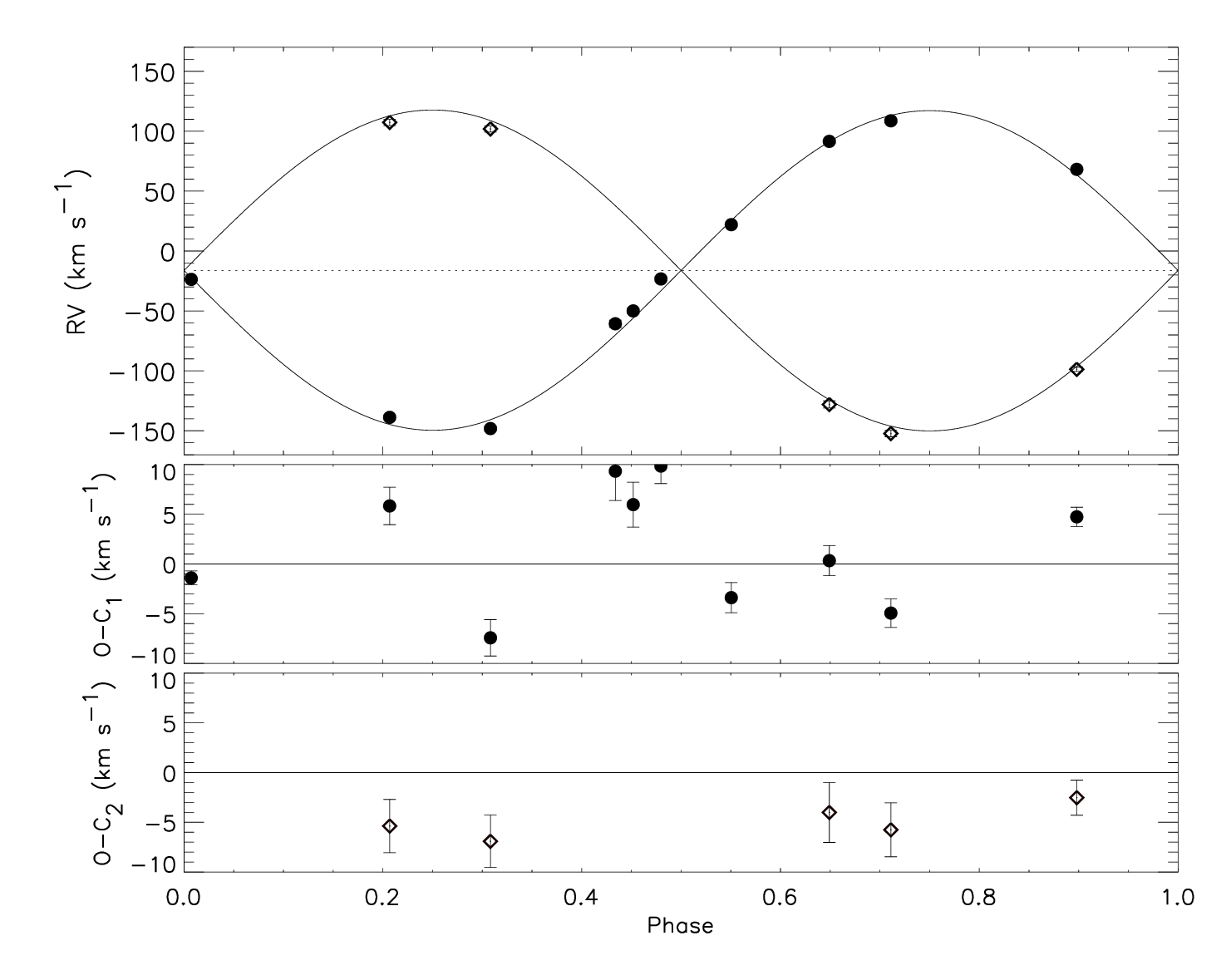}
  \centering
  \includegraphics[width=.45\linewidth]{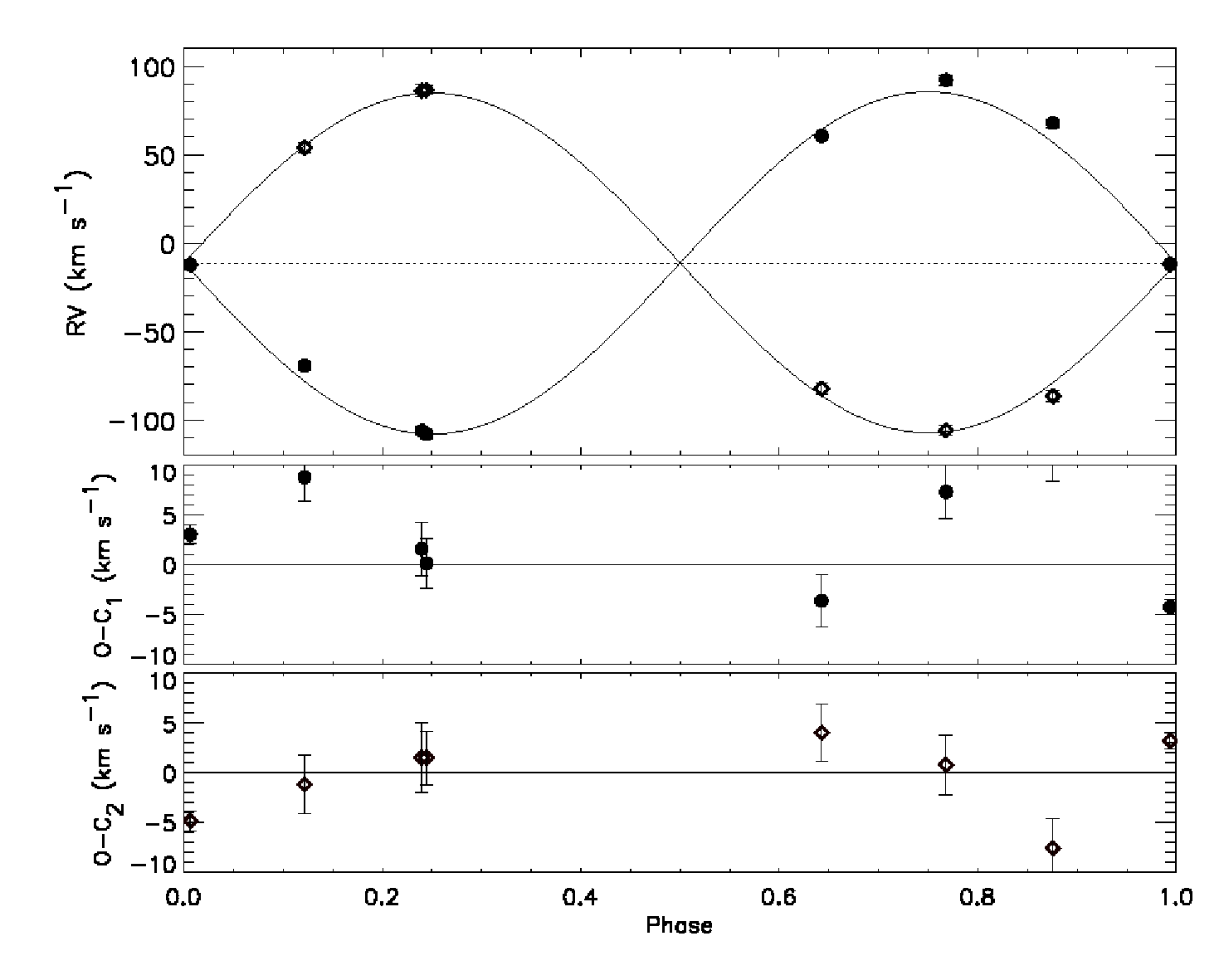}
\end{subfigure}
\caption{Comparison of the theoretical radial velocity curve with the observations for HD\,117476 and 205\,Dra (upper panels). The middle and bottom panels represent the residuals between the observation and theoretical RV variation for primary (star showing eclipse at deep minimum) and secondary (star showing eclipse at deep minimum) stars (see Table\,\ref{tab:lc} also), respectively.}
\label{fig:rv1}
\end{figure*}

The RV measurements were subsequently analyzed utilizing the \texttt{rvfit} program\footnote{\url{http://www.cefca.es/people/~riglesias/rvfit.html}}, which models RV data of double-lined, single-lined binaries, or exoplanets using the minimization method of adaptive simulated annealing \citep{2015PASP..127..567I}. During this analysis, one can search for the orbital parameters of systems such as orbital period (P$_{orb}$), periastron passage time (T$_{0}$), eccentricity ($e$), the argument of the periastron ($\omega$), RV of the center of mass ($\gamma$), projected semi-major axis ($a\sin i$, where $i$ is orbital inclination) and the semi-amplitude of RV for the primary\footnote{star showing eclipse at deep minimum} (p) and secondary\footnote{star showing eclipse at deep minimum} (s) (K$_{p,s}$). In the RV analysis we fixed the systems's orbital periods given in Table\,\ref{table1} and searched for the other parameters. Our RV measurements were used in this study. However, for the HY\,Vir and V1031\,Ori case the literature RV measurements \citep{2011AJ....142..185S, 1990A&A...228..365A} were also collected and used in our RV analysis in addition to our measurements. 

\begin{figure*}
\centering
\begin{subfigure}{1\textwidth}
  \centering
  \includegraphics[width=.45\linewidth]{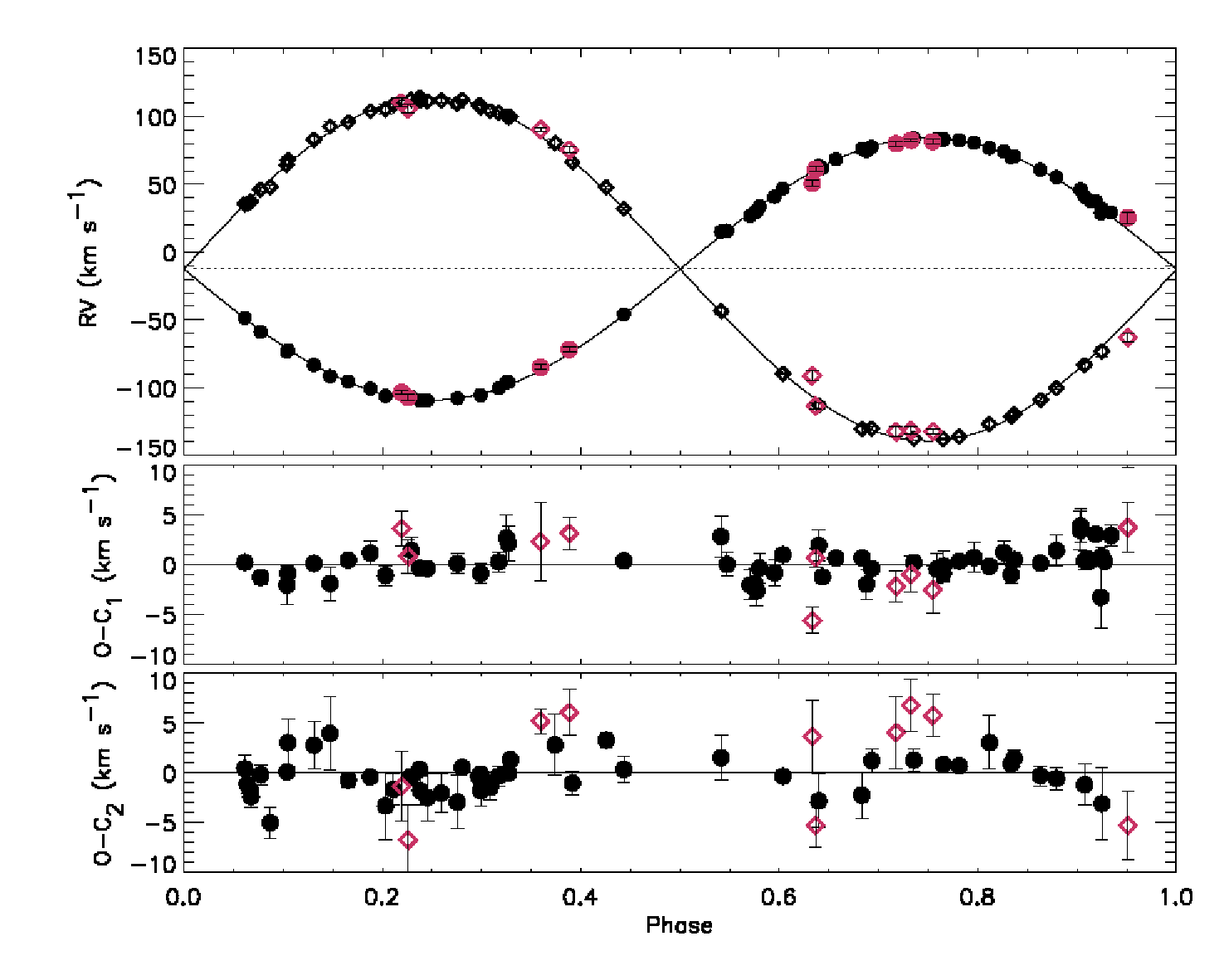}
  \centering
  \includegraphics[width=.45\linewidth]{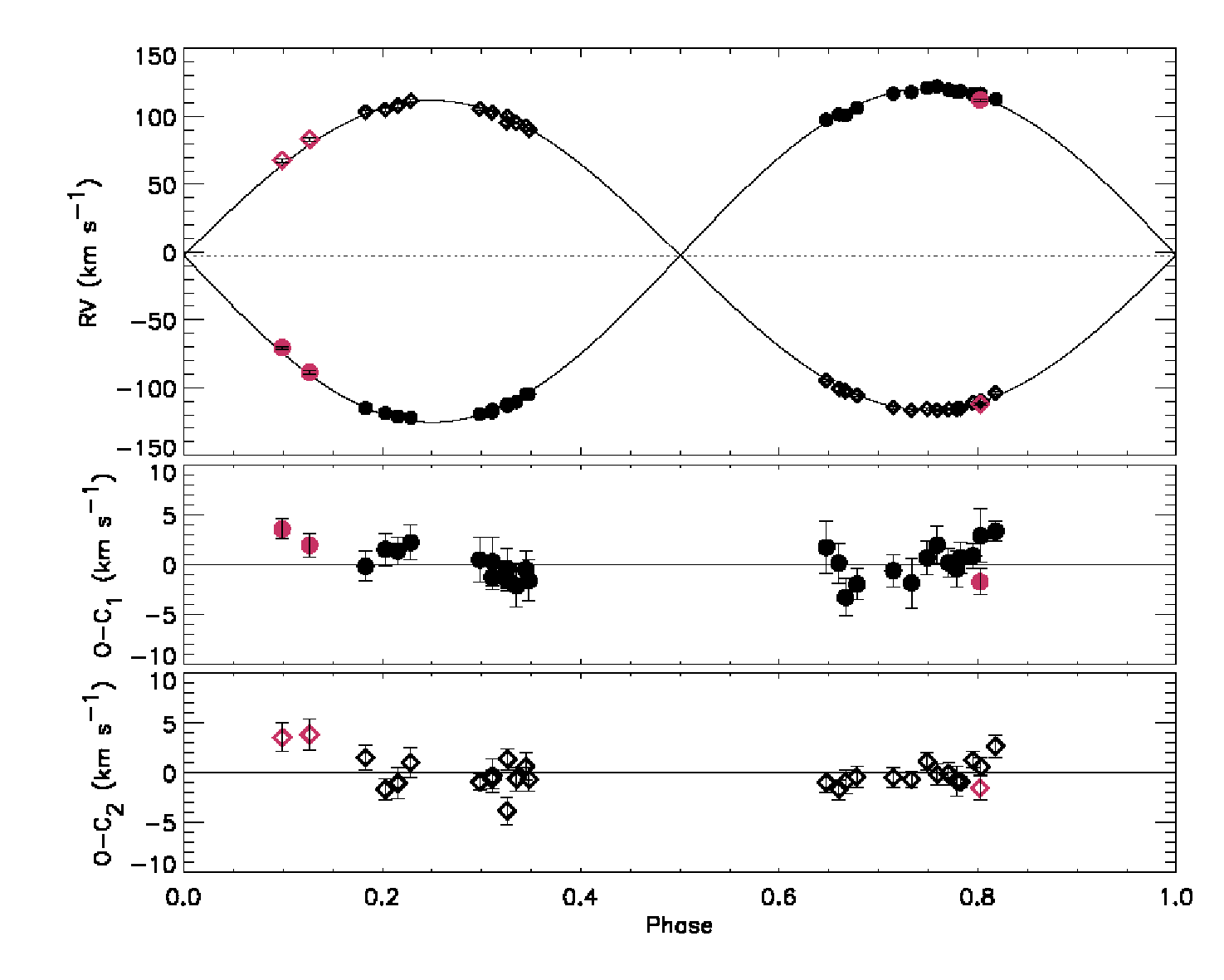}
\end{subfigure}
\caption{Comparison of the theoretical radial velocity curve with the observation for HY\,Vir and V1031\,Ori (upper panels). The red diamond symbols are the observational data, while the other black symbols are the literature data from \cite{2011AJ....142..185S} for HY\,Vir, and \cite{1990A&A...228..365A} for V1031\,Ori. The middle and bottom panels represent the residuals between the observation and theoretical RV variation for primary (star showing eclipse at deep minimum) and secondary (star showing eclipse at deep minimum) stars, respectively.}
\label{fig:rv2}
\end{figure*}

The analysis, initially, aimed to determine whether the systems exhibit eccentric orbits. Subsequently, it was established that all targets possess circular orbits. For this reason, the $e$ and $\omega$ values were fixed at 0 and 90 degrees, respectively, for further analysis. As a result, the orbital parameters of the systems were determined and they are listed in Table\,\ref{rvresult}. The consistency between the measured RV values and the theoretical RV curves is shown in Fig.\,\ref{fig:rv1} and Fig.\,\ref{fig:rv2}. V1031\,Ori was found to be a triple system in the study of \cite{1990A&A...228..365A} and they found the radial velocity shift of the third component. However, since the variation of the third component does not affect the analysis of the double system we excluded these data from our analysis. In addition, we refer to the results of \cite{2021PASJ...73..809L}, who reported that the third component of V1031\,Ori has a projected separation of approximately 90\,mas and an orbital period of about 92 years. These parameters indicate that the third component is widely separated from the inner binary, and its influence on the measured radial velocities over the timescale of our observations is negligible.

\begin{table*}
\begin{center}
\centering
\caption[]{The results of the RV analysis. The subscripts p and s represent the primary (star showing eclipse at deep minimum) and the secondary (star showing eclipse at deep minimum) components (see Table\,\ref{tab:lc} also), respectively. $^a$ shows the fixed parameters. The errors are given in parentheses in units of the last decimal. $f(m_p,m_s)$ is the mass function.}\label{rvresult}
\begin{tabular*}{0.7\linewidth}{@{\extracolsep{\fill}}lrrrr}
\hline
 Parameter		  &  HD\,117476 	    & 205\,Dra      & HY\,Vir       & V1031\,Ori         \\
\hline
$T_0$ (HJD+2450000)     & 8000.427\,(8) &2003.222\,(3)  &4000.671\,(1)   &14000.013\,(1)   \\
$\gamma$ (km/s)	         & -17.6\,(4)	 &-11.2\,(4)     &-12.3\,(1)      &-2.5\,(2) \\
$K_p$ (km/s)	         & 132.9\,(8)    &96.8\,(9)      &96.4\,(1)       &123.1\,(3) \\
$K_s$ (km/s)	         & 133.3\,(1.3)  &96.1\,(9)      &125.8\,(1)      &114.3\,(2) \\
$a_p\sin i$ ($R_\odot$)	 & 3.45\,(2)     &8.11\,(10)     &5.21\,(1)       &8.28\,(2) \\
$a_s\sin i$ ($R_\odot$)	 & 3.46\,(3)     &8.06\,(11)     &6.81\,(1)       &7.69\,(2) \\
$a  \sin i$ ($R_\odot$)	 & 6.91\,(4)     &16.17\,(15)    &12.02\,(1)      &15.97\,(3)\\
$M_p\sin ^3i$ ($M_\odot$)& 1.29\,(3)	 &1.57\,(5)      &1.76\,(1)       &2.27\,(1)\\
$M_s\sin ^3i$ ($M_\odot$)& 1.28\,(2)     &1.58\,(5)      &1.35\,(1)       &2.45\,(1)\\
$q = M_s/M_p$	         & 0.997\,(1)    &1.007\,(20)    &0.766\,(1)      &1.077\,(3)\\
 \hline
\end{tabular*}
     \end{center}
\end{table*}

\subsection{Determination of atmospheric parameters}

After estimating the orbital parameters through RV analysis, we conducted another spectroscopic analysis to estimate the atmospheric parameters (e.g., \teff, surface gravity, \logg) of each binary component. Since our targets include both double and triple systems (V1031 Ori), their spectra contain flux contributions from two or even three stellar components, depending on the system. Therefore, analysis methods for single stars are not applicable. Two methods are commonly used for the analysis of such systems. The first is known as spectral disentangling \citep{1994A&A...281..286S}. With the spectral disentangling method, the spectra of the binary component stars can be separated from the observed composite spectrum by considering the flux ratio of the binary components. The other method involves creating a composite synthetic spectrum that represents the observed spectrum, incorporating the flux from both binary components. In this method, the flux ratio between the components of binary system is an important quantity for proper renormalization of disentangled spectra \citep{2010ASPC..435..207P}. The second method can be applied in any system which have spectra showing the lines of binary components. However, for proper disentangling the observed spectra should be distributed over the orbital cycle. Therefore, by examining the RV curves of the target systems, we decided to apply the disentangling method to HD\,117476, 205\,Dra, and HY\,Vir, as we have their spectra 
distributed across the orbital phase (see Figs.,\ref{fig:rv1} and \ref{fig:rv2}). In contrast, we only have three spectra for V1031\,Ori, so the second method was applied to this system instead.

\begin{figure*}
 \centering
 \begin{minipage}[b]{0.45\textwidth}
  \includegraphics[height=6.5cm, width=1\textwidth]{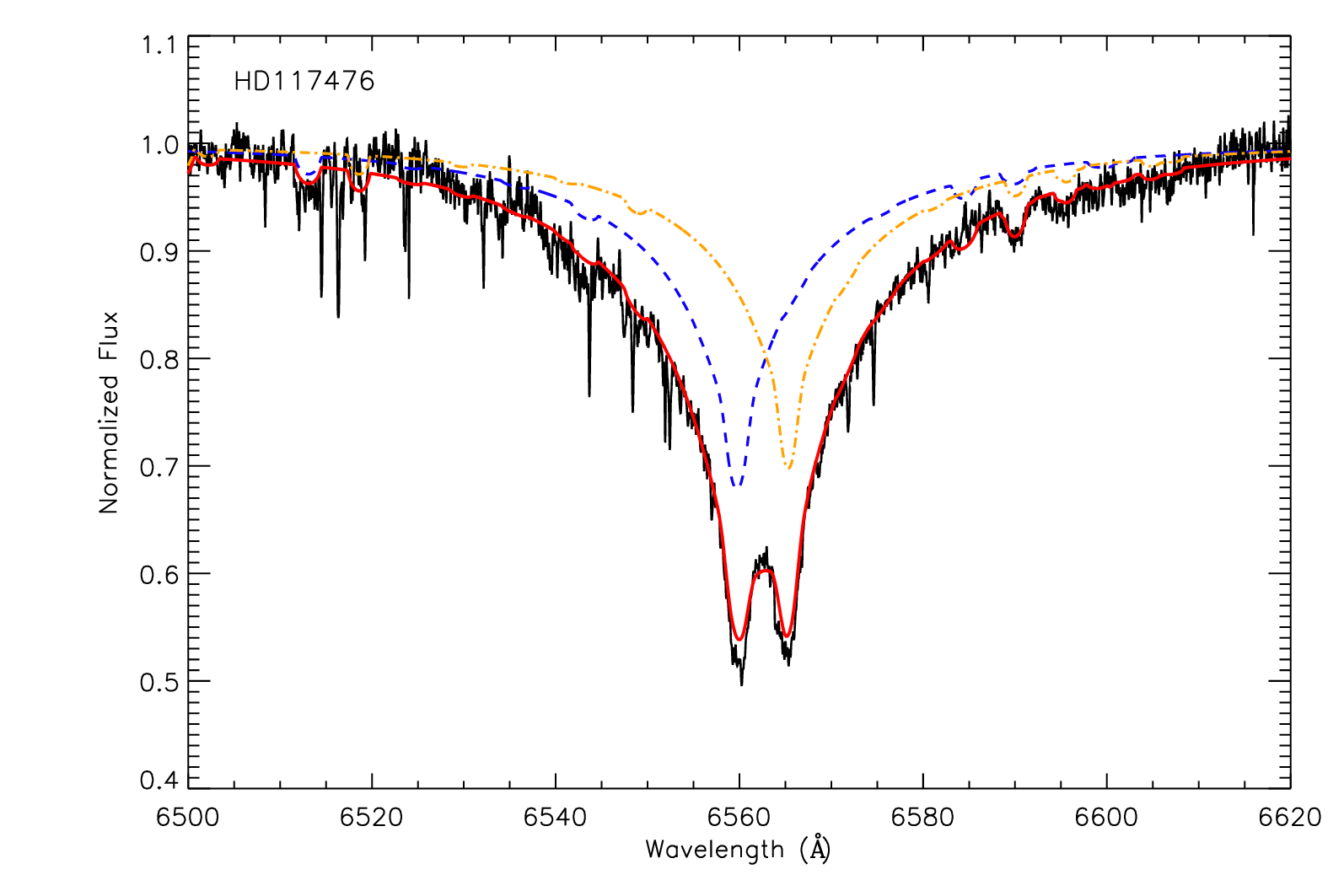}
 \end{minipage}
 \begin{minipage}[b]{0.45\textwidth}
  \includegraphics[height=6.5cm, width=1\textwidth]{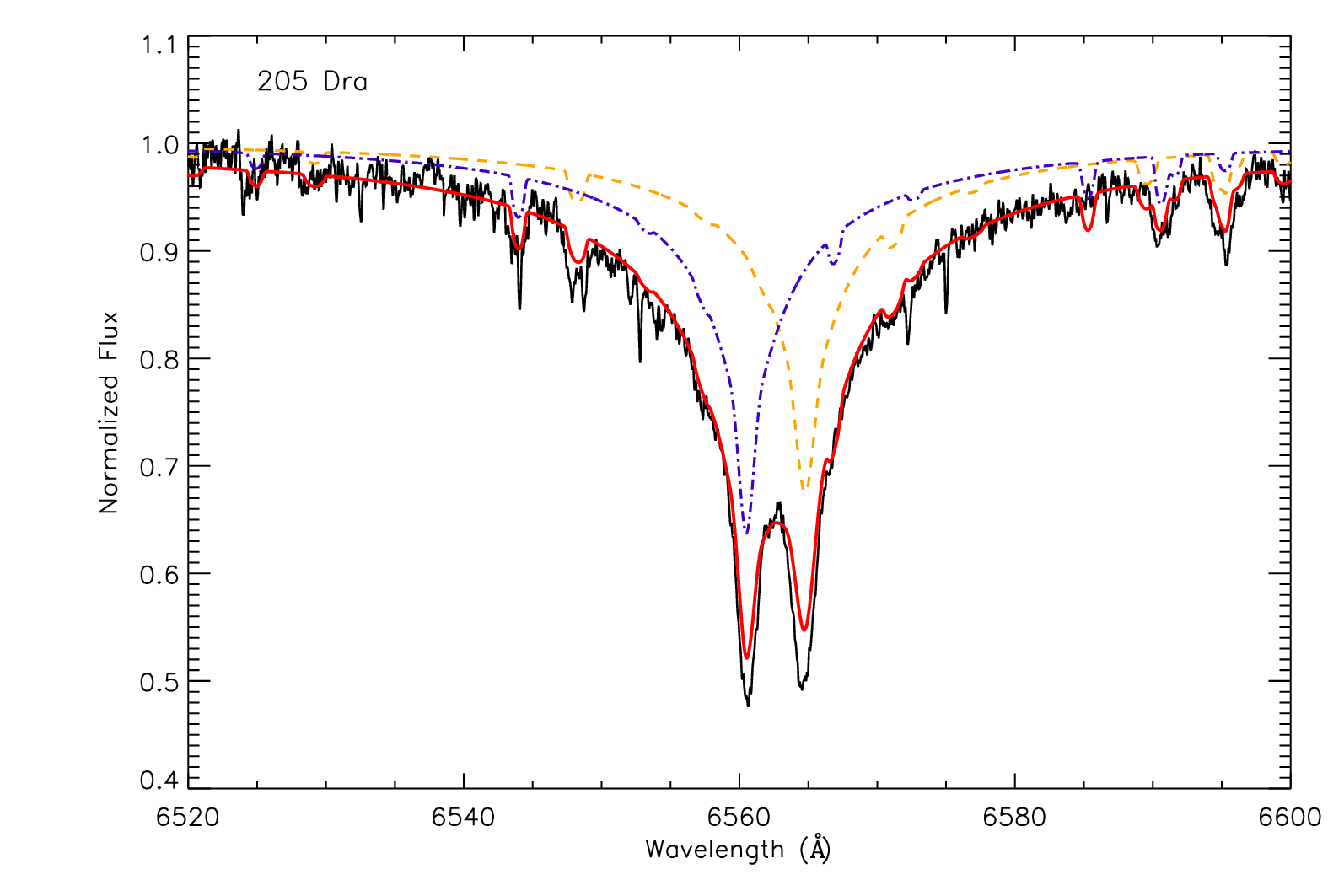}
  \end{minipage}
   \begin{minipage}[b]{0.45\textwidth}
  \includegraphics[height=6.5cm, width=1\textwidth]{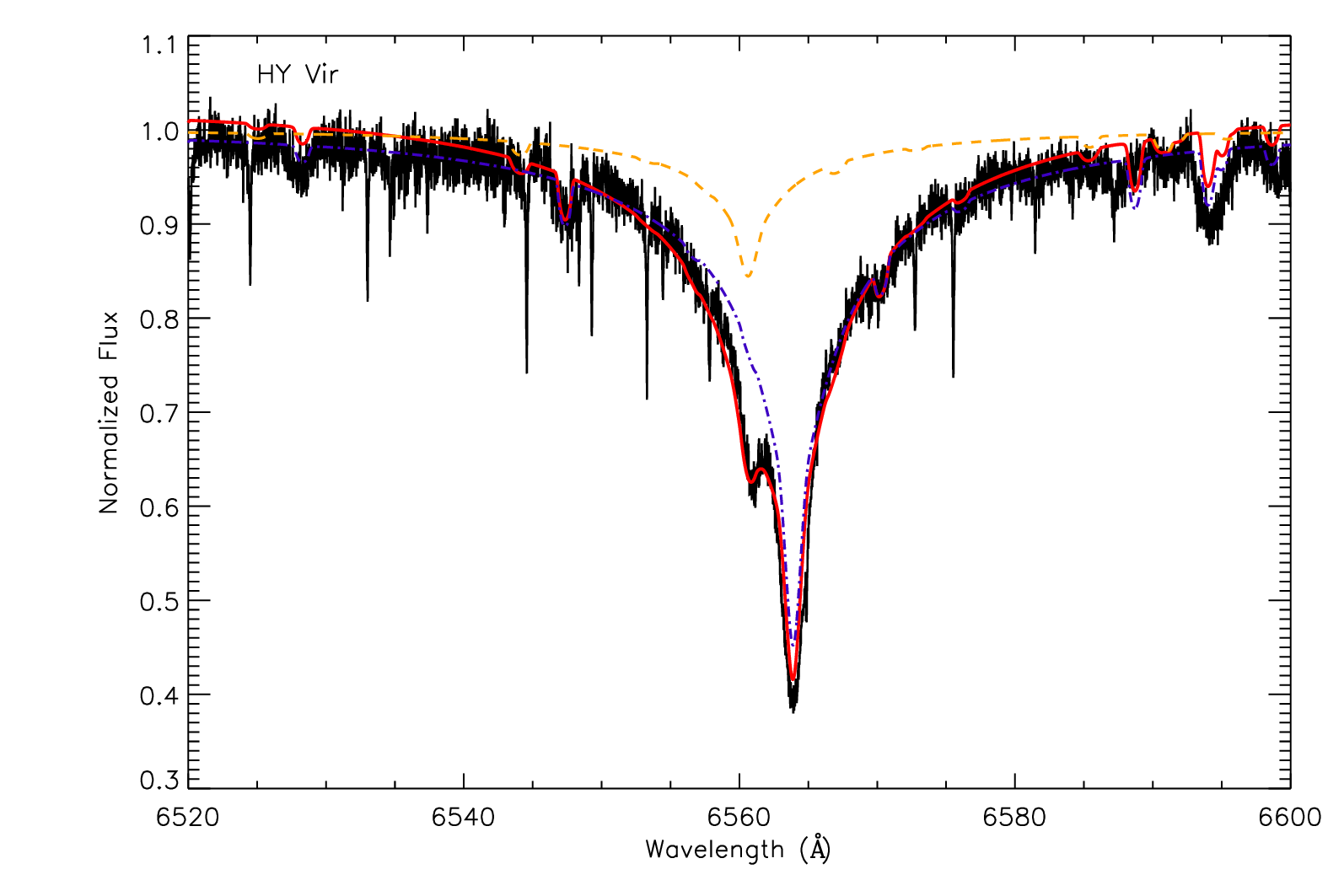}
  \end{minipage}
   \begin{minipage}[b]{0.45\textwidth}
  \includegraphics[height=6.5cm, width=1\textwidth]{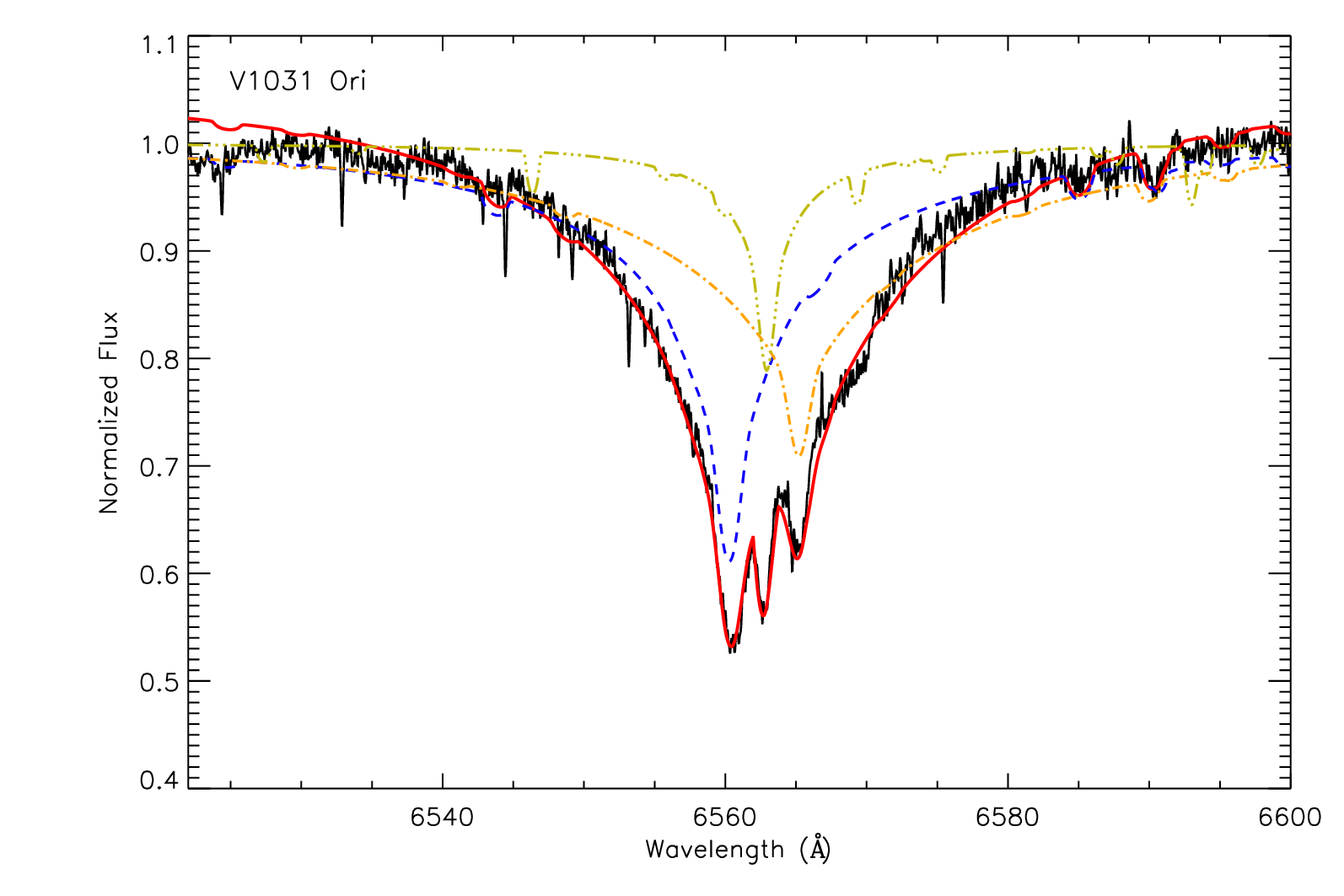}
  \end{minipage}
   \caption{The theoretical composite spectra (red line) fit the observed H$_\alpha$ lines (black lines). The blue and yellow lines represent the theoretical spectra of the primary (star showing eclipse at deep minimum) and secondary (star showing eclipse at deep minimum) components (see Table\,\ref{tab:lc} also) of each eclipsing binary system. The light green line in the V1031\,Ori panel displays the synthetic spectra of third component.}  \label{hfits}
\end{figure*}

To implement the disentangling method, we used the \texttt{FDBINARY} code \citep{2004ASPC..318..111I}, which disentangles a composite spectrum in Fourier space and can also resolve the spectrum of a third component. Prior to applying the method, we performed initial binary modeling to estimate the flux ratios of the binary components for all systems (for details, see Sec.,4.1). These flux ratios were then used as input for the \texttt{FDBINARY} code and to create a composite spectrum for V1031\,Ori. For the \texttt{FDBINARY} analysis, the flux ratios, along with the orbital parameters determined from the RV analysis, were used as input. As a result of \texttt{FDBINARY} application, the spectra of each binary component were derived and re-normalized using the method outlined by \cite{2004ASPC..318..111I}.

\begin{table*}
\centering
  \caption{The H$\alpha$ \teff\, values, the final atmospheric parameters, \vsini\, and the Fe abundances of the component stars of HD\,117476, 205\,Dra, and HY\,Vir. The subscripts p and s represent the primary (hooter star showing eclipse at deep minimum) and the secondary (cooler star showing eclipse at deep minimum) components (see Table\,\ref{tab:lc} also), respectively.}
  \label{tab:atmospar}
  \begin{tabular*}{0.9\linewidth}{@{\extracolsep{\fill}}lcccccc}
\hline
                 &\multicolumn{1}{c}{\hrulefill H$\alpha$ line\,\hrulefill}
                 &\multicolumn{5}{c}{\hrulefill \,Fe lines\,\hrulefill}\\
Star             & \teff\,\,(K)   &  \teff\,\,(K)     & \logg\,\,(cgs)     & $\xi$\,(\kms)   & \vsini\,\,(\kms) & $\log \epsilon$ (Fe)  \\
\hline
HD\,117476$_{p}$   &8000\,$\pm$\,200   &7800\,$\pm$\,100   & 4.0\,$\pm$\,0.1    & 2.30\,$\pm$\,0.2 & 66\,$\pm$\,4&7.38\,$\pm$\,0.32\\
HD\,117476$_{s}$   &7800\,$\pm$\,200   &7900\,$\pm$\,200   & 4.3\,$\pm$\,0.1    & 3.00\,$\pm$\,0.2 & 62\,$\pm$\,4&8.21\,$\pm$\,0.34\\
205\,Dra$_{p}$     &7000\,$\pm$\,200   &7000\,$\pm$\,200 & 3.7\,$\pm$\,0.2  & 3.5\,$\pm$\,0.2 & 38\,$\pm$\,4 &8.24\,$\pm$\,0.32\\
205\,Dra$_{s}$     &6900\,$\pm$\,200   &6800\,$\pm$\,100 & 3.5\,$\pm$\,0.1  & 3.0\,$\pm$\,0.2 & 39\,$\pm$\,3 &7.50\,$\pm$\,0.35\\
HY\,Vir$_{p}$      &7100\,$\pm$\,200   &7300\,$\pm$\,200 & 3.8\,$\pm$\,0.1  & 2.5\,$\pm$\,0.2 & 50\,$\pm$\,1 &8.17\,$\pm$\,0.33\\
HY\,Vir$_{s}$      &6900\,$\pm$\,200   &7100\,$\pm$\,100 & 4.2\,$\pm$\,0.1  &1.6\,$\pm$\,0.2  & 33\,$\pm$\,1 &7.58\,$\pm$\,0.37\\
\hline
\end{tabular*}
\end{table*}

The spectrum of each binary component was individually analyzed using the ATLAS9 model atmosphere and SYNTHE code. Initially, the hydrogen Balmer lines were considered to estimate the effective temperature (\teff) values following the method outlined by \cite{2004A&A...425..641C}. Subsequently, the metal line diagnostic was employed to refine and determine the values of \teff, surface gravity (\logg), microturbulence velocity ($\xi$), and the projected rotational velocity (\vsini) \citep{2014dapb.book.....N, 2016MNRAS.458.2307K}. In this diagnostic analysis, iron (Fe) element lines were utilized due to their prevalence within the effective temperature range of our target stars, as outlined in \cite{2005oasp.book.....G, 2014dapb.book.....N, 2016MNRAS.458.2307K}. To estimate the atmospheric parameters of V1031\,Ori we utilized the second method; creating composite spectrum. Since V1031\,Ori is a triple system with all components visible in our spectra, the flux contributions of the binary components were determined from preliminary binary modeling and taken into account during the analysis. We generated numerous synthetic spectra based on the parameters provided by \cite{1990A&A...228..365A}, using the ATLAS9 and SYNTHE codes. These synthetic spectra were computed for various \teff\, values in the range 7000–9500\,K (with steps of 100\,K), and \logg\, values from 3.5 to 4.5 (in steps of 0.1), and composite spectra were constructed by considering the flux contributions of each component. For the synthetic spectra of the third component, \logg\, and metallicity were fixed to 4.0 and solar, respectively, due to its less flux contribution to the composite spectrum (see next section). As its relatively lower flux contribution made it difficult to constrain these parameters through hydrogen line analysis. The resulting composite spectra were then compared with the observed spectra to determine the final atmospheric parameters of the components of V1031\,Ori.

The derived atmospheric parameters are given in Table\,\ref{tab:atmospar} for the systems analyzed with the disentangling method. For V1031\,Ori the determined atmospheric parameters are given in Table\,\ref{tab:atmosparv1031}. The consistency between the observed and theoretical models is shown in Fig.\,\ref{hfits} by creating composite spectra considering the derived atmospheric parameters of the binary components to give the same figure for all targets. 

\begin{table}
\centering
  \caption{The \teff, \logg, and \vsini\, values for V1031\,Ori.
  The subscripts p, s and t represent the primary (hotter star showing eclipse at deep minimum), the secondary (cooler star showing eclipse at deep minimum), and third  components (see Table\,\ref{tab:lc} also), respectively. * displays the fixed parameters. The \vsini\, value for the third component was taken from \cite{1990A&A...228..365A}.}
  \label{tab:atmosparv1031}
  \begin{tabular}{lccc}
\hline
Star             & \teff\,\,(K)   &  \logg\,\,(cgs)     &  \vsini\,\,(\kms)   \\
\hline
V1031\,Ori$_{p}$   &8500\,$\pm$\,200   & 3.9\,$\pm$\,0.2    & 45\,$\pm$\,4\\
V1031\,Ori$_{s}$   &7600\,$\pm$\,200   & 3.6\,$\pm$\,0.2    & 62\,$\pm$\,5\\
V1031\,Ori$_{t}$   &8600\,$\pm$\,500   & 4.0*    & 9*\\
\hline
\end{tabular}
\end{table}

\begin{figure*}
 \centering
 \begin{minipage}[b]{0.45\textwidth}
  \includegraphics[height=4.5cm, width=1\textwidth]{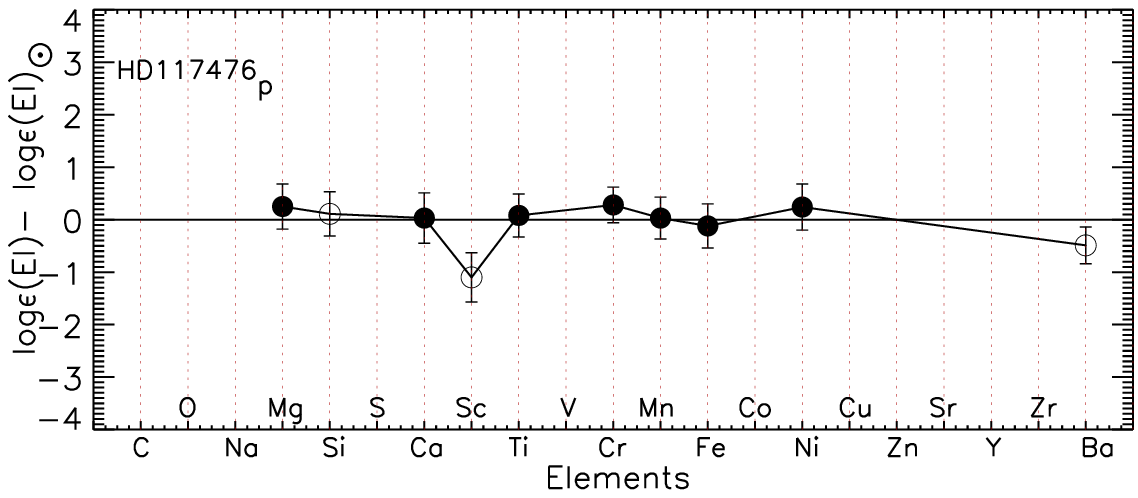}
 \end{minipage}
 \begin{minipage}[b]{0.45\textwidth}
  \includegraphics[height=4.5cm, width=1\textwidth]{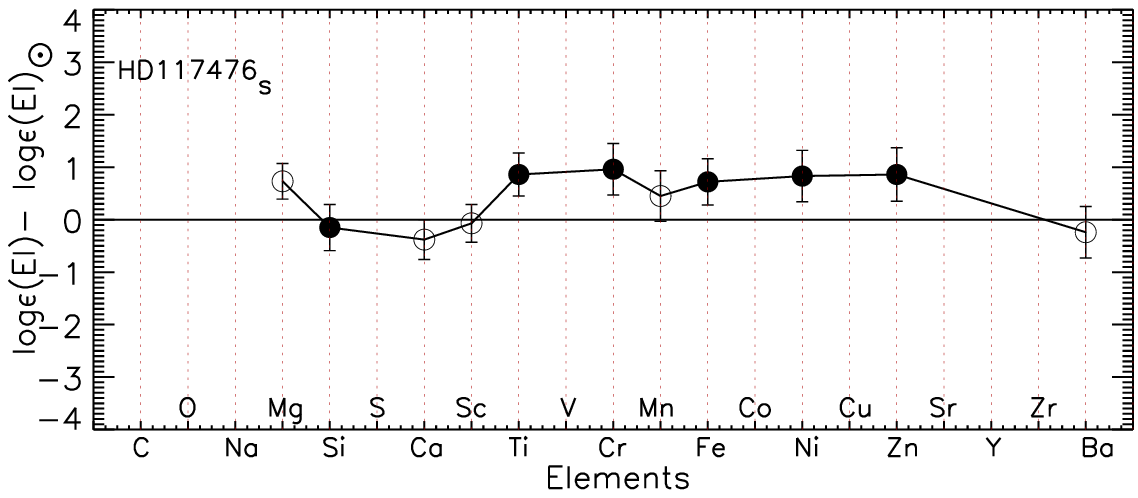}
  \end{minipage}
   \begin{minipage}[b]{0.45\textwidth}
  \includegraphics[height=4.5cm, width=1\textwidth]{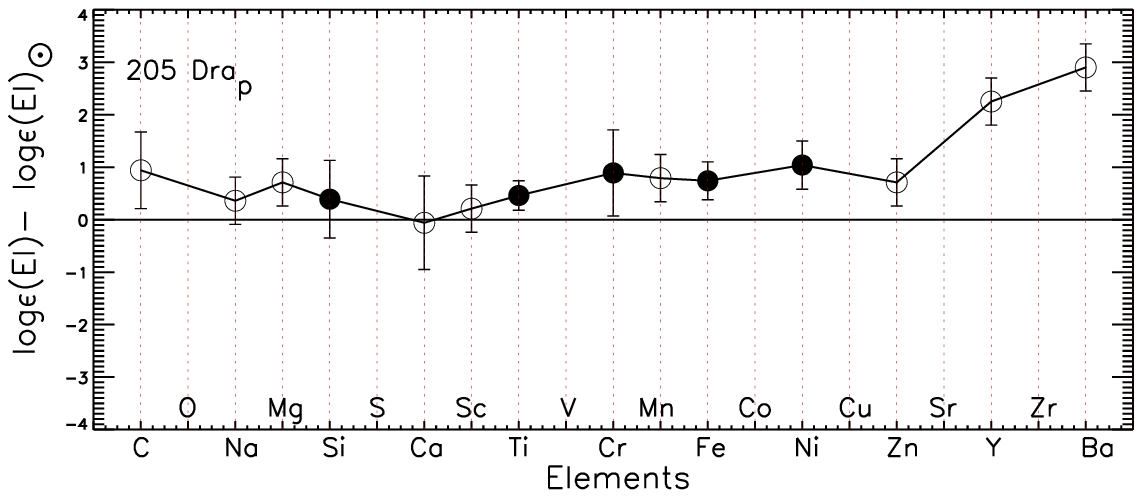}
  \end{minipage}
   \begin{minipage}[b]{0.45\textwidth}
  \includegraphics[height=4.5cm, width=1\textwidth]{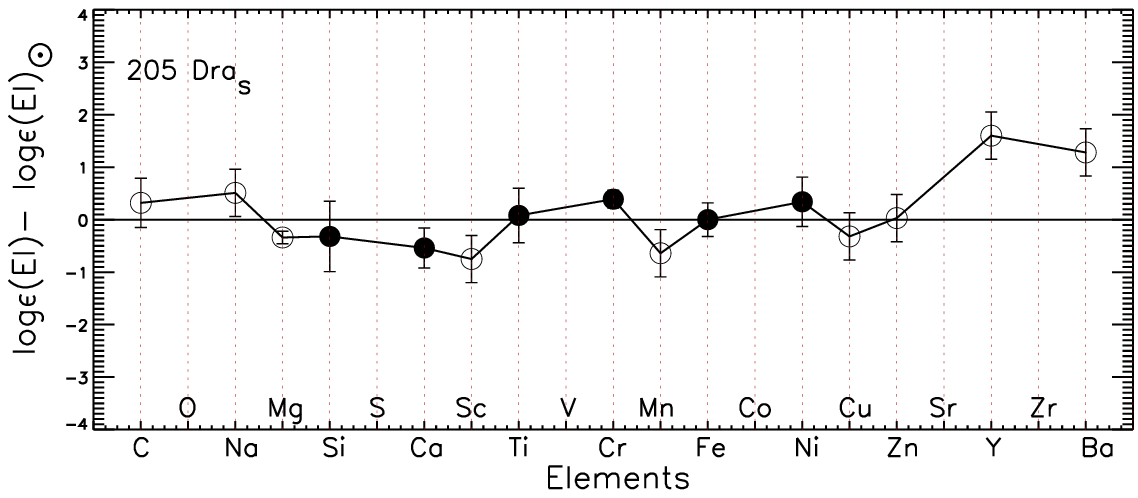}
  \end{minipage}
    \begin{minipage}[b]{0.45\textwidth}
  \includegraphics[height=4.5cm, width=1\textwidth]{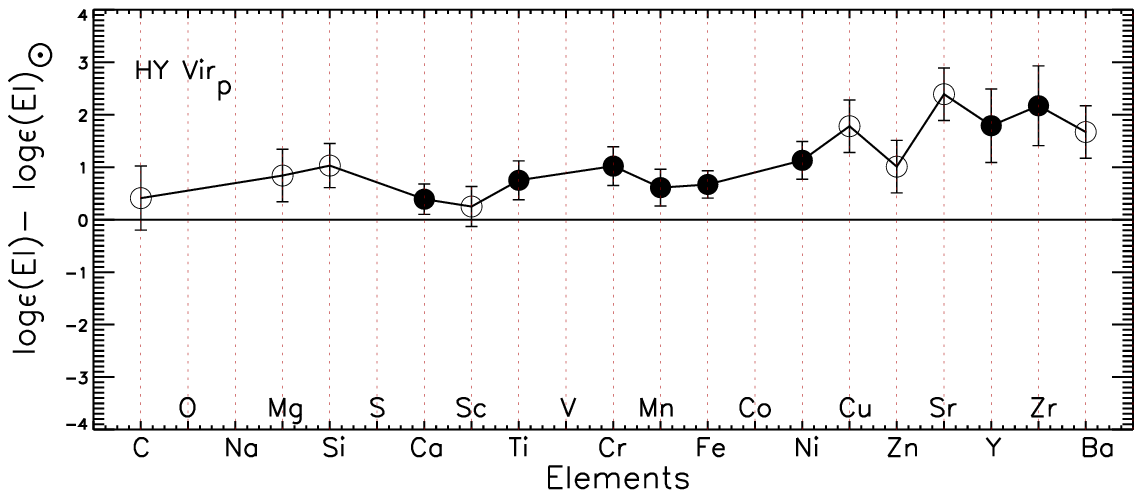}
  \end{minipage}
    \begin{minipage}[b]{0.45\textwidth}
  \includegraphics[height=4.5cm, width=1\textwidth]{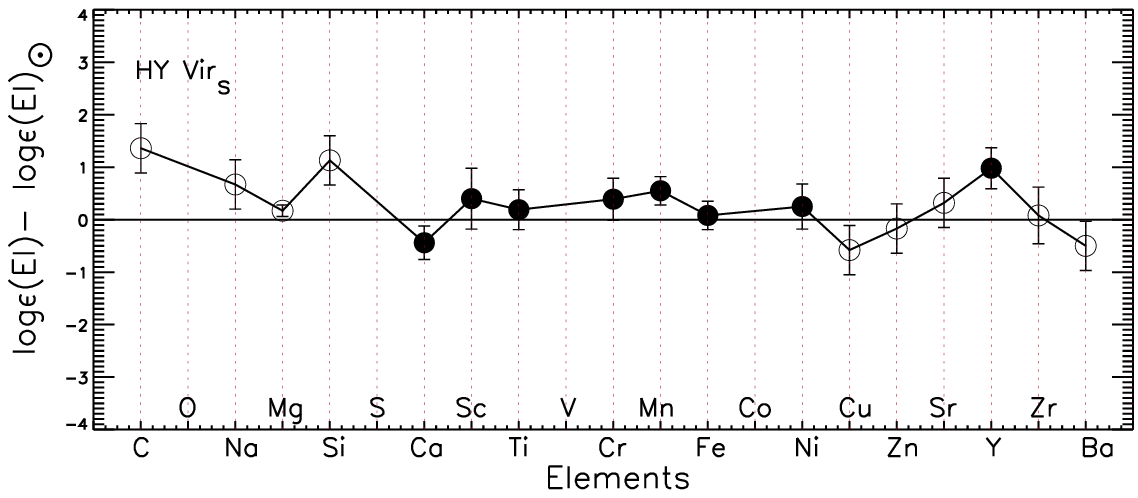}
  \end{minipage}
   \caption{Chemical abundance distributions of the binary components relative to solar abundance \citep{2009ARA&A..47..481A}. The subscripts ``p'' and ``s'' represent the primary and the secondary components, respectively. Abundances from five or more lines are shown with filled symbols, while open symbols indicate fewer than five lines.}  \label{fig:abundance}
\end{figure*}

After estimating the final atmospheric parameters listed we derived the chemical abundances of individual elements for the systems which analyzed with the disentangling method. The spectrum synthesis method was preferred for the chemical abundance analysis, as it is more suitable for moderate to high rotating stars ($\gtrapprox$40\,\kms) \citep{2014dapb.book.....N}. The same model atmospheres and synthesis code were used throughout the analysis. Before the analysis, line identification was performed using the Kurucz atomic line list\footnote{kurucz.harvard.edu/linelists.html} (gfhyperall.dat), which includes wavelengths, excitation potentials, and $\log gf$ values. The spectrum synthesis method takes into account the given atmospheric parameters and the list of elements responsible for the absorption lines. By varying the abundances of these elements and minimizing the difference between the observed and synthetic spectra, the method determines the optimal elemental abundances. We calculated the abundances of several elements for the binary components of HD\,117476, 205\,Dra, and HY\,Vir. The determined chemical abundances for the components of target binary systems are listed in Table\,\ref{tab:A2}, and their distributions are shown in Fig.\,\ref{fig:abundance}. The Fe element abundances are also provided in Table\,\ref{tab:atmospar}. The uncertainties in the derived absolute abundances were estimated by considering the typical errors in the input atmospheric parameters, following the methodology described in \citet{2016MNRAS.458.2307K}. To assess the sensitivity of the abundance results we systematically varied \teff\, \logg\,, $\xi$, and \vsini. Additionally, the effects of spectral resolution and signal-to-noise ratio on the abundance determinations were taken into account based on the same reference. All these sources of uncertainty were combined in quadrature to provide the final error estimates listed for the absolute abundances. The absolute Fe abundances of some binary components listed in Table\,\ref{tab:atmospar} were found to be higher than the solar Fe abundance \citep[7.50;][]{2009ARA&A..47..481A}. As shown in Fig.\,\ref{fig:abundance}, the primary components of 205\,Dra and HY\,Vir, as well as—though to a lesser extent—the secondary component of HD\,117476, appear to exhibit an Am star–like abundance pattern \citep[e.g.,][]{2007BaltA..16..183A, 2009A&A...503..945F, 2010A&A...523A..71G}.

\section{Photometric analysis}

\begin{figure*}
 \centering
 \begin{minipage}[b]{0.45\textwidth}
  \includegraphics[height=6.5cm, width=1\textwidth]{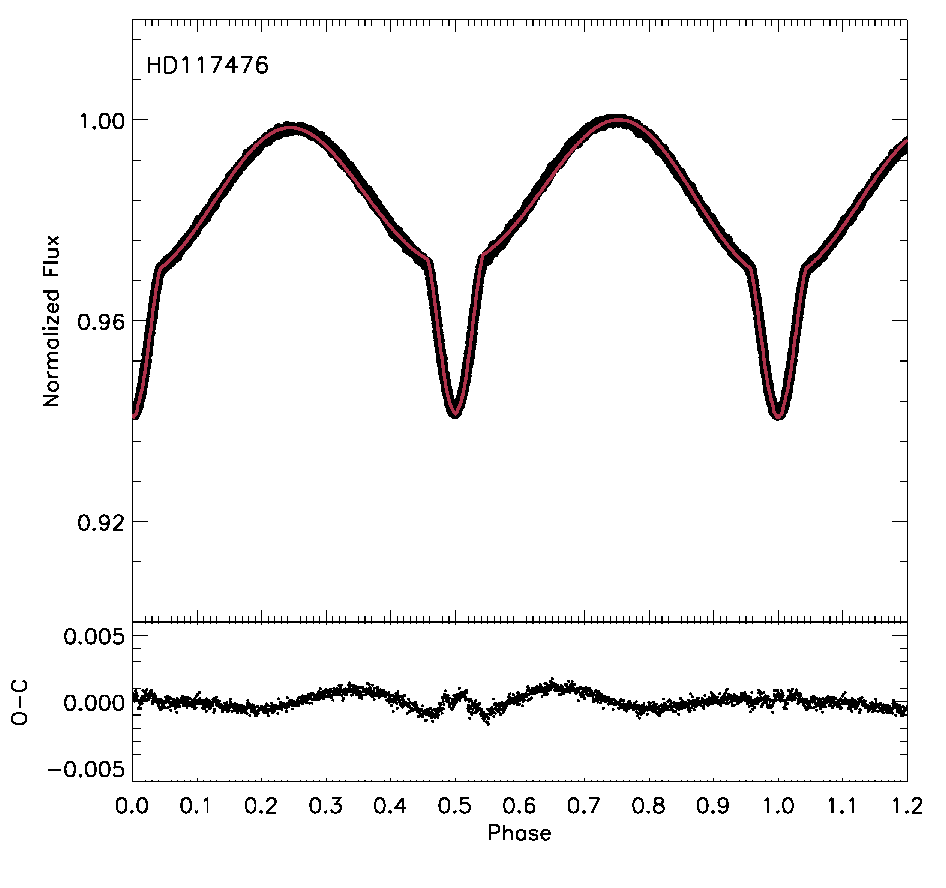}
 \end{minipage}
 \begin{minipage}[b]{0.45\textwidth}
  \includegraphics[height=6.5cm, width=1\textwidth]{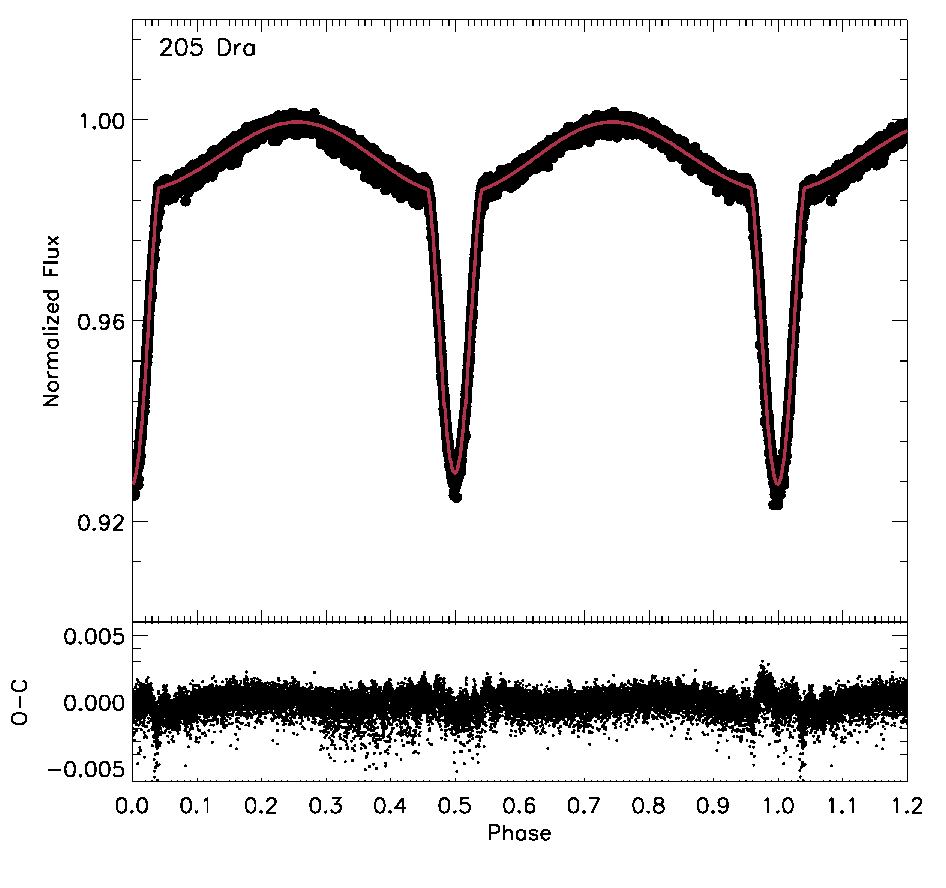}
  \end{minipage}
   \begin{minipage}[b]{0.45\textwidth}
  \includegraphics[height=6.5cm, width=1\textwidth]{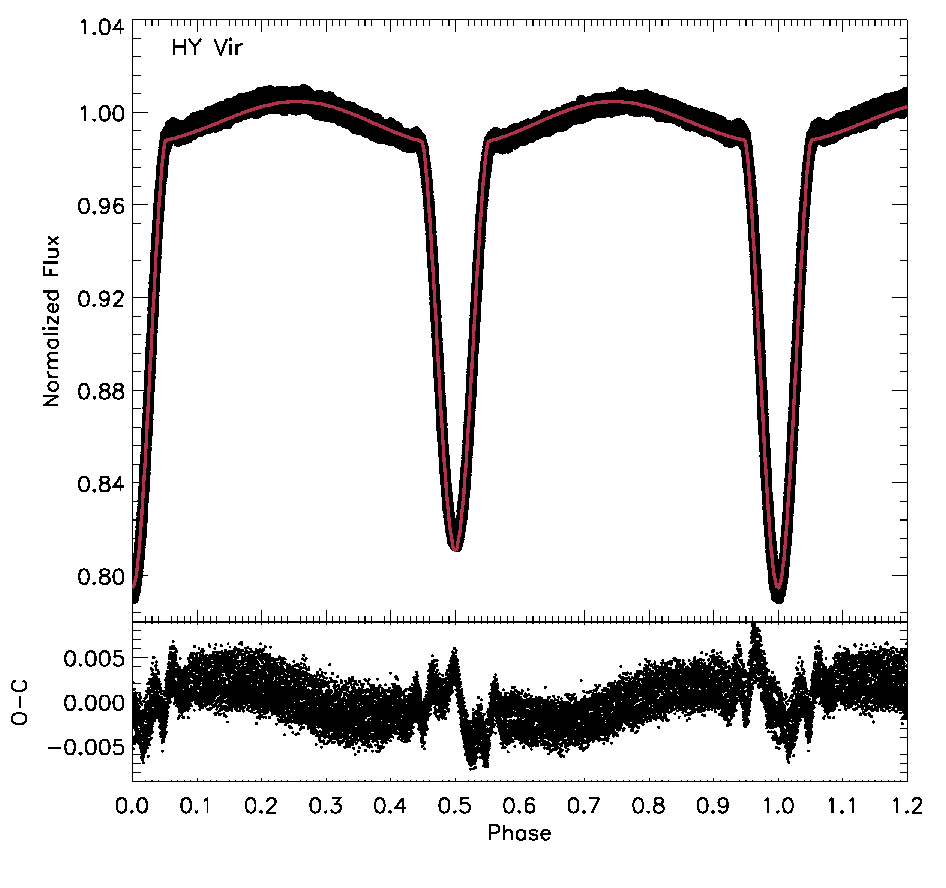}
  \end{minipage}
   \begin{minipage}[b]{0.45\textwidth}
  \includegraphics[height=6.5cm, width=1\textwidth]{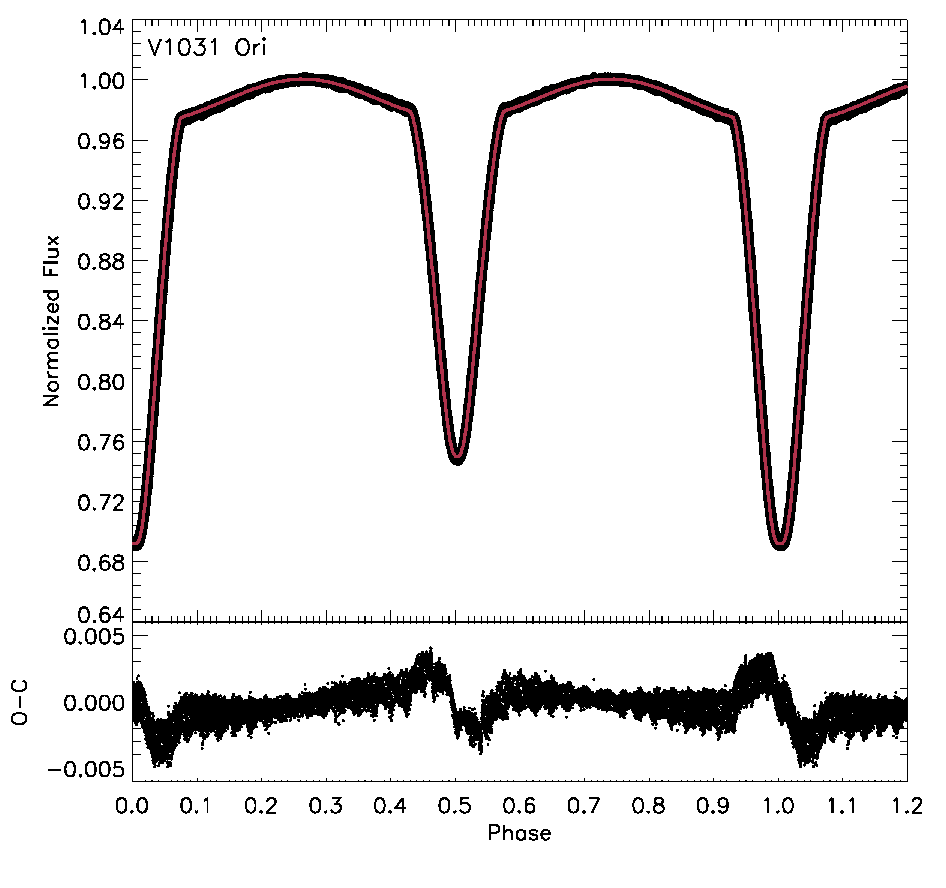}
  \end{minipage}
   \caption{Upper panels: The theoretical binary models (red solid lines) fit to the TESS photometric data of the target systems. Lower panels: residuals.}  \label{lcfits}
\end{figure*}

\subsection{Binary modeling}

All TESS photometric data of the targets were used for binary modeling. Before analysis, these TESS data were examined, and if the eclipse depths were affected by oscillations, the pulsations were first subtracted from the light curves. For this purpose, the pulsations were initially modeled using the \textsc{Period04} program \citep{2005CoAst.146...53L}, which applies the Fourier transform to the time series of the photometric observations. The modeled pulsations were then subtracted from the entire light curve of the selected systems. The resulting binary light curves were binned to 4000 points to make them usable by the binary modeling code.

For the analysis of binarity, we used the Wilson-Devinney binary modeling code \citep[W-D,][]{1971ApJ...166..605W}, combined with Monte Carlo simulations for the calculation of the uncertainties \citep{2004AcA....54..299Z, 2010MNRAS.408..464Z}. During this analysis, the \teff\, values of the primary binary components (the hotter stars) were taken from our spectroscopic analysis and kept fixed throughout the investigation. In addition to the \teff\, values of the primary components, other parameters were also fixed, including the bolometric albedos \citep{1969AcA....19..245R}, the bolometric gravity-darkening coefficient \citep{1924MNRAS..84..665V}, the logarithmic limb darkening coefficient \citep{1993AJ....106.2096V} as specified in \cite{2022MNRAS.510.1413K}. In our modelling, we adopted bolometric albedo values of 1.0 for stars with radiative envelopes and 0.5 for stars with convective envelopes. Similarly, the gravity-darkening coefficients were set to 1.0 for radiative envelopes following \citet{1924MNRAS..84..665V} law and 0.32 for convective envelopes as suggested by \citet{1967ZA.....65...89L}.The mass ratios ($q$). The following parameters are optimized: the dimensionless potential ($\Omega$), the \teff\, of the secondary component, inclination ($i$), phase shift ($\phi$), and the relative luminosities of the binary components. Initially, to estimate the binary configuration of the systems, many iterations were performed, and the $\Omega$ values were monitored to determine whether the binary components filled their Roche lobes. Consequently, it was found that the target systems are detached binaries, and the detached binary configuration was considered in further binary modeling. The best-fitting binary models were determined from minimum of the sum of squared values and the final models were obtained. The results of the binary modelings are given in Table\,\ref{tab:lc}, and the best fitting binary models' consistency is shown in Fig.\,\ref{lcfits}. 

\begin{table*}
\begin{center}
\centering
\caption{Results of the binary modeling and the astrophysical parameters.
The subscripts p, s, and 3 represent the primary (star showing eclipse at deep minimum), secondary (star showing eclipse at deep minimum), and third components. $^a$ and $l$ show the fixed parameters and relative luminosity, respectively.}\label{tab:lc}
\begin{tabular}{lrrrr}
\hline
 Parameter			             & HD\,117476   	      &205\,Dra            &HY\,Vir         & V1031\,Ori\\	
\hline
$i$ ($^{o}$)	       	          & 65.17 $\pm$ 0.01       &72.45 $\pm$ 0.01   &79.74 $\pm$ 0.01 &85.02 $\pm$ 0.02    \\	
\teff$_{p}$$^a$ (K)               & 7800 $\pm$ 200  	   &7100 $\pm$ 200	  &7300 $\pm$ 200   &8500 $\pm$ 200 \\	
\teff$_{s}$ (K)    	              & 7890 $\pm$ 225	       &6900 $\pm$ 100	  &7100 $\pm$ 230	&7800 $\pm$ 220    	\\
$\Omega$$_{p}$		              & 5.050 $\pm$ 0.010      &6.754 $\pm$ 0.017  &7.607 $\pm$ 0.016 &6.736 $\pm$ 0.004	    	  \\
$\Omega$$_{s}$		              & 5.092 $\pm$ 0.015      &5.622 $\pm$ 0.002  &4.682 $\pm$ 0.005 &4.983 $\pm$ 0.001 	     	  \\
$\phi$              	          & 0.0009 $\pm$ 0.0001    &-0.0009 $\pm$ 0.0001 &-0.0004 $\pm$ 0.0001 &0.0026 $\pm$ 0.0001    \\
$q$                               & 0.997 $\pm$ 0.001      &1.007 $\pm$ 0.019  	&0.766 $\pm$ 0.001 & 1.077 $\pm$ 0.003    	  \\
$r$$_{p}$$^*$ (mean)              & 0.249 $\pm$ 0.002      &0.174 $\pm$ 0.002   &0.216 $\pm$ 0.001 & 0.177 $\pm$ 0.001   \\
$r$$_{s}$$^*$ (mean)              & 0.246 $\pm$ 0.002      &0.217 $\pm$ 0.001   &0.146 $\pm$ 0.001 & 0.271 $\pm$ 0.001   \\
$l$$_{p}$ / ($l$$_{p}$+$l$$_{s}$) & 0.499 $\pm$ 0.02        &0.39 $\pm$ 0.02  	&0.68 $\pm$ 0.02   & 0.35 $\pm$ 0.01 	  \\
$l$$_{s}$ /($l$$_{p}$+$l$$_{s}$)  & 0.501 $\pm$ 0.02        &0.61 $\pm$ 0.02 	&0.32 $\pm$ 0.01   & 0.65 $\pm$ 0.02 	  \\
$l$$_{3}$                         &       -                 & -	                 & -               & 0.21\\
\multicolumn{4}{c}{Derived Quantities}\\

$M$$_{p}$ ($M_\odot$)	          & 1.72 $\pm$ 0.03  	   &1.81 $\pm$ 0.05		 &1.86 $\pm$ 0.01 &2.30 $\pm$ 0.01   \\	
$M$$_{s}$ ($M_\odot$)	          & 1.72 $\pm$ 0.02        &1.82 $\pm$ 0.05		 &1.43 $\pm$ 0.04 &2.48 $\pm$ 0.01   	\\
$R$$_{p}$ ($R_\odot$)	          & 1.90 $\pm$ 0.07        &2.93 $\pm$ 0.03		 &2.70 $\pm$ 0.01 &2.84 $\pm$ 0.05    	  \\
$R$$_{s}$ ($R_\odot$)		      & 1.87 $\pm$ 0.06        &3.73 $\pm$ 0.04		 &1.79 $\pm$ 0.06 &4.34 $\pm$ 0.07  	  \\
log\,$L$$_{p}$ ($L_\odot$)        & 1.08 $\pm$ 0.02        &1.29 $\pm$ 0.03		 &1.27 $\pm$ 0.02 &1.58 $\pm$ 0.04  	  \\
log\,$L$$_{s}$ ($L_\odot$)	      & 1.09 $\pm$ 0.01        &1.45 $\pm$ 0.03      &0.86 $\pm$ 0.03 &1.80 $\pm$ 0.04  	  \\
\logg\,$_{p}$ (cgs)               & 4.12 $\pm$ 0.07 	   &3.76 $\pm$ 0.06  	 &3.84 $\pm$ 0.05 &3.88 $\pm$ 0.04    			    \\
\logg\,$_{s}$ (cgs)               & 4.13 $\pm$ 0.09 	   &3.56 $\pm$ 0.01		 &4.08 $\pm$ 0.06 &3.55 $\pm$ 0.03    			    \\
$M_{bol}$$_{p}$ (mag)      & 2.05 $\pm$ 0.05        & 1.62 $\pm$ 0.07         	 &1.56 $\pm$ 0.06 &0.79 $\pm$ 0.01 \\
$M_{bol}$$_{s}$ (mag)	  & 2.03 $\pm$ 0.07        &1.12 $\pm$ 0.07	         	 &2.59 $\pm$ 0.05 &0.24 $\pm$ 0.01 \\
$M_{V}$$_{p}$ (mag)	              & 1.97 $\pm$ 0.09        &1.58 $\pm$ 0.07	     &1.49 $\pm$ 0.09 &0.82 $\pm$ 0.02   	  \\
$M_{V}$$_{s}$ (mag)	              & 2.38 $\pm$ 0.18        &1.08 $\pm$ 0.07      &2.50 $\pm$ 0.10 &0.20 $\pm$ 0.04   		  \\
Distance (pc)                     & 181  $\pm$ 7             &210 $\pm$ 6        & 202 $\pm$ 6    & 179$\pm$ 4       \\
 \hline
\end{tabular}
     \end{center}
     \begin{description}
     \centering
 \item[ ] *fractional radius, $R$/$a$.
 \end{description}
\end{table*}

The uncertainties of the provided parameters were estimated using Monte Carlo simulations and by varying parameters such as albedo and gravity darkening with different values (e.g., Southworth 2020; Pavlovski et al. 2023). The derived uncertainties of the parameters are given in Table\,\ref{tab:lc} as well. The resulting parameters from the binary modeling were used to calculate fundamental stellar parameters such as mass ($M$), radius ($R$), luminosity ($L$), bolometric ($M_{bol}$), and absolute ($M_{V}$) magnitudes. To compute these parameters, the interstellar reddening, $E(B-V)$, the Kepler and Pogson equations, along with the bolometric correction (BC) from the study by Eker et al. (2020), were utilized. The $E(B-V)$ values were estimated from the interstellar extinction map of \citet{2005AJ....130..659A} and found to be 0.009$\pm$0.002, 0.016$\pm$0.002, 0.010$\pm$0.002, and 0.024$\pm$0.002 mag for HD\,17476, 205\,Dra, HY\,Vir, and V1031\,Ori, respectively. These parameters are also provided in Table\,\ref{tab:lc}.

\begin{table*}
\centering
\caption{The pulsation frequencies of targets. Formal error estimates are given in brackets in units of the last digits after the comma.}
\begin{tabular}{rccc|rccc}
\hline
&  & \multicolumn{2}{c}{HD117476}  & &  & \multicolumn{2}{c}{205\,Dra}  \\
\hline
& \multicolumn{1}{c}{Frequency} & \multicolumn{1}{c}{Amplitude} & \multicolumn{1}{c}{SNR} & & \multicolumn{1}{c}{Frequency} & \multicolumn{1}{c}{Amplitude} & \multicolumn{1}{c}{SNR} \\
& \multicolumn{1}{c}{d$^{-1}$} & \multicolumn{1}{c}{mmag} & \multicolumn{1}{c}{} & & \multicolumn{1}{c}{d$^{-1}$} & \multicolumn{1}{c}{mmag} & \multicolumn{1}{c}{}   \\
& & \multicolumn{1}{c}{$\pm 0.02$} &  & & & \multicolumn{1}{c}{$\pm 0.02$} &  \\
\hline
$f_1$\,+\,2$f_{orb}$ & 27.5097 (3)& 0.11 &  10 & $f_1$ & 11.8873 (1)& 2.53 &  38  \\
$f_1$ & 25.9878 (1) & 0.34 &  37 & $f_1$\,-\,3$f_{orb}$ & 11.1791 (2)& 0.39 &  27\\
$f_2$\,+\,$f_{orb}$ & 32.3377 (3) & 0.11 &  10 & $f_2$ & 9.4946 (1)& 2.29 &  33 \\
$f_2$ & 31.5772 (1) & 0.27 &  33 & $f_2$\,-\,2$f_{orb}$ & 9.0221 (1)& 1.70 &  6\\
$f_2$\,-\,$f_{orb}$ & 30.8158 (1) & 0.24 &  29 & $f_3$ & 11.4149 (1)& 1.96 &  28\\
$f_3$ & 31.3355 (1) & 0.23 &  27 & $f_4$ & 11.0304(1)& 1.42 &  19\\
$f_3$\,-\,$f_{orb}$ & 30.5740 (4) & 0.02 &  8 & $f_5$ & 12.6425 (2)& 1.37 &  22\\
$f_3$\,-\,2$f_{orb}$ & 29.8136 (2)  & 0.15 &  15 & $f_6$ & 13.1717 (2)& 1.30 &  24\\
$f_3$\,-\,4$f_{orb}$ & 28.2897 (4) & 0.06 &  6 & $f_6$\,-\,2$f_{orb}$ & 12.6993 (3)& 0.35 &  6\\
$f_4$\,+\,2$f_{orb}$ & 27.7798 (4) & 0.09 &  8 & $f_7$ & 13.0387 (2)& 1.22 &  22\\
$f_4$ & 26.2560 (2) & 0.22 &  21 & $f_8$ & 12.8753 (2)& 1.05 &  18\\
$f_4$\,-\,2$f_{orb}$ & 24.7350 (3) & 0.09 &  13 & $f_8$\,-\,2$f_{orb}$ & 12.4038 (3)& 0.60 &  10 \\
$f_5$\,+\,2$f_{orb}$ & 27.1945(4) & 0.08 &  7 &$f_9$ & 8.9820 (2)& 0.93 &  14\\
$f_5$\,+\,$f_{orb}$ & 26.4351 (3)  & 0.14 &  13 & $f_10$=$f_1$+$f_6$-$f_3$ & 13.1150 (2)& 0.82 &  15\\
$f_5$ & 25.6717 (3) & 0.17 &  20 &$f_11$ & 11.5049 (2)& 0.82 &  12\\
$f_6$\,+\,4$f_{orb}$ & 31.7015 (4)& 0.06 &  7 &$f_12$ & 12.5701 (2)& 0.69 &  12\\
$f_6$\,+\,2$f_{orb}$ & 30.1796 (3)& 0.10 &  10 &$f_13$ & 11.6937 (2)& 0.69 &  12\\
$f_{6}$& 28.6616 (3) & 0.12 &  12 & $f_14$=$f_2$+$f_8$-$f_12$ & 11.2182 (2)& 0.68 &  11\\
$f_{7}$ & 28.0118 (3) & 0.09 &  8 &$f_15$ & 13.2236 (2)& 0.69 &  12\\
$f_{8}$ & 25.5885 (3) & 0.07 &  8 &$f_16$=$f_1$+$f_14$-$f_3$ & 11.9773 (2)& 0.69 &  9 \\
\hline
&  & \multicolumn{2}{c}{HY\,Vir}  & &  & \multicolumn{2}{c}{V1031\,Ori}  \\
\hline
& \multicolumn{1}{c}{Frequency} & \multicolumn{1}{c}{Amplitude} & \multicolumn{1}{c}{SNR} & & \multicolumn{1}{c}{Frequency} & \multicolumn{1}{c}{Amplitude} & \multicolumn{1}{c}{SNR} \\
& \multicolumn{1}{c}{d$^{-1}$} & \multicolumn{1}{c}{mmag} & \multicolumn{1}{c}{} & & \multicolumn{1}{c}{d$^{-1}$} & \multicolumn{1}{c}{mmag} & \multicolumn{1}{c}{}   \\
& & \multicolumn{1}{c}{$\pm 0.02$} &  & & & \multicolumn{1}{c}{$\pm 0.02$} &  \\
\hline
$f_1$ & 13.4101 (1)& 0.77 &  45 & $f_1$ & 11.4926 (1)& 0.80 &  70  \\
$f_2$ & 12.1768 (1) & 0.40 &  21 & $f_2$ & 8.8086 (2)& 0.60 &  46\\
$f_3$ & 11.8872 (1) & 0.37 &  12 & $f_3$ & 11.7867 (2)& 0.20 &  16 \\
$f_4$ & 10.2954 (1) & 0.36 &  10 & $f_4$ & 12.3750 (2)& 0.13 &  9\\
$f_5$ & 12.6162 (1) & 0.36 &  13 & $f_5$ & 12.6711 (2)& 0.10 &  8\\
$f_6$ & 11.4420 (2) & 0.13 &  6  & $f_6$ & 10.6179 (1)& 0.07 &  5\\
$f_7$ & 7.2390 (2)  & 0.11 &  5 &  & &  &  \\
$f_8$ & 14.0149 (2) & 0.12 &  7 &  & &  & \\
$f_9$ & 10.7289 (2) & 0.11 &  11 & & &  & \\
\hline
\end{tabular}
\label{tab1:puls_freq}
\end{table*}

\subsection{Pulsation analysis}

The pulsational structures of the systems were revealed by our independent investigations and also with the studies of \cite{2022RAA....22h5003K, 2023MNRAS.524..619K}. In the current pulsational analysis, we carried out a comprehensive examination to understand the detailed pulsational behavior of the systems. The 120-s SAP flux TESS data were used to analyze the binary-variation–removed light curves for pulsational frequency analysis. For this purpose, we utilized the same method described by \cite{2023MNRAS.524..619K}. The harmonics and multiples of the orbital frequencies of the orbital periods were fitted to the whole light curves, and then these fits were removed. Consequently, only the pulsational variations of the binary components were obtained.

\begin{figure}
  \centering
  \includegraphics[height=5cm, width=0.5\linewidth]{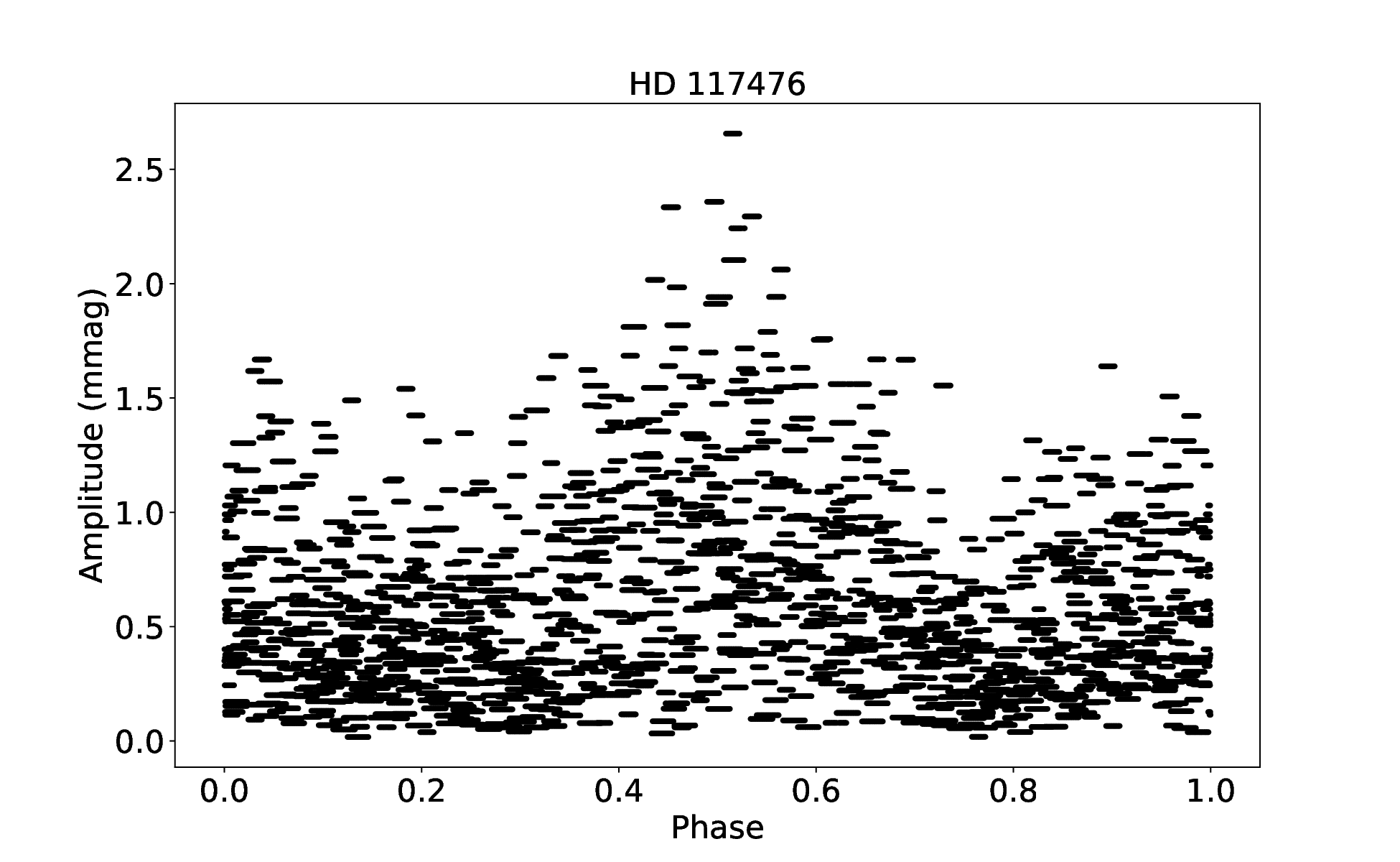}  
\caption{Amplitude change of the highest amplitude frequency of HD\,117476 with the orbital phase.}
\label{fig:4}
\end{figure}

The pulsation frequency analysis of the systems was carried out with the \textsc{Period04} program \citep{2005CoAst.146...53L}. The lists of the determined frequencies are given in Table\,\ref{tab1:puls_freq} and the theoretical pulsation fit to the observations are shown in the Appendix.  We took into account the study given by \cite{2021AcA....71..113B} to determine the significance limit for the frequencies. As a result, frequencies were identified for the systems as given in the table. In the case of HD\,117476 we found that some frequencies are combinations of certain frequencies with the orbital period's frequency and its multiples. This frequency distribution is commonly observed in tidally tilted pulsators \citep[e.g.][]{2020NatAs...4..684H, 2022MNRAS.510.1413K, 2021MNRAS.503..254R}. The pulsation amplitude in these tidally tilted pulsators varies with the orbital phase. Therefore, we investigated how the amplitude of the highest amplitude frequency changes with the orbital phase in the case of HD\,117476. The amplitude variation of this frequency according to the orbital phase is shown in Fig. \ref{fig:4}. As seen in the figure, the amplitude reaches its maximum and minimum values at around phases 0.5 and 0.25-0.75, respectively. This is an expected variation for a tidally tilted pulsator. Since the frequencies of the system show modulation with the orbital period (see Appendix Fig.\,\ref{fig2:appen}), we classified HD\,117476 as a candidate tidally tilted pulsator.


According to the determined astrophysical parameters (particularly $M$, $L$ and \teff) both components of each binary system are located within the \ds\ instability strip (see Sect.\ \ref{sect:evo}). Therefore, it is challenging to identify which component(s) in these binary systems exhibits oscillations. To address this, we followed a few approaches.

In the first approach, we calculated the expected frequencies for the radial fundamental mode and its overtones, based on the pulsation constant values ($Q$) from \cite{1981ApJ...249..218F}.  The fundamental parameters of each binary component were considered and the expected frequencies were calculated using the equation $Q\,=\,P\sqrt{\rho/\rho_{\odot}}\,=\,P M^{1/2} R^{-3/2}$. For the primary and secondary components of HD\,117476, the expected frequency ranges are 14\,–\,27\,d$^{-1}$ and 19\,–\,27\,d$^{-1}$, respectively. However, since the determined frequencies fall within these estimated ranges, it is difficult to distinguish the pulsating component in HD\,117476. For 205\,Dra, the calculated frequency ranges are 8\,–\,16\,d$^{-1}$ for the primary component and 6\,–\,11\,d$^{-1}$ for the secondary component. Since the observed frequencies for 205\,Dra mostly fall within the primary component’s frequency range, we assumed the primary component is the pulsating one according to this approach. For HY\,Vir, the calculated frequencies are in the range of 17\,–\,34\,d$^{-1}$ for the primary and 8\,–\,16\,d$^{-1}$ for the secondary component. Based on this information and the determined frequency ranges for HY\,Vir, we inferred that the secondary component is likely the oscillating one. In the case of V1031\,Ori, the system consists of three components, all of which appear to be potential pulsator based on their \teff\, and \logg\, values. Therefore, the expected frequencies were calculated separately for each component. For the third component of V1031\,Ori, the $M$ and $R$ values were estimated using the parameters listed in Table\,\ref{tab:atmosparv1031} and the \teff\,-$M$ and \teff\,-$R$ calibration tables from \citet{2000asqu.book.....C}. After examining the calculated oscillation frequencies and the expected frequency ranges for its components (for the primary: 10\,–\,19\,d$^{-1}$; for the secondary: 5\,–\,10\,d$^{-1}$, for the third component: 15\,–\,23\,d$^{-1}$), we concluded that the primary component is most likely the pulsating star according to this approach. 

\begin{figure*}
 \centering
 \begin{minipage}[b]{0.45\textwidth}
  \includegraphics[height=4.5cm, width=1\textwidth]{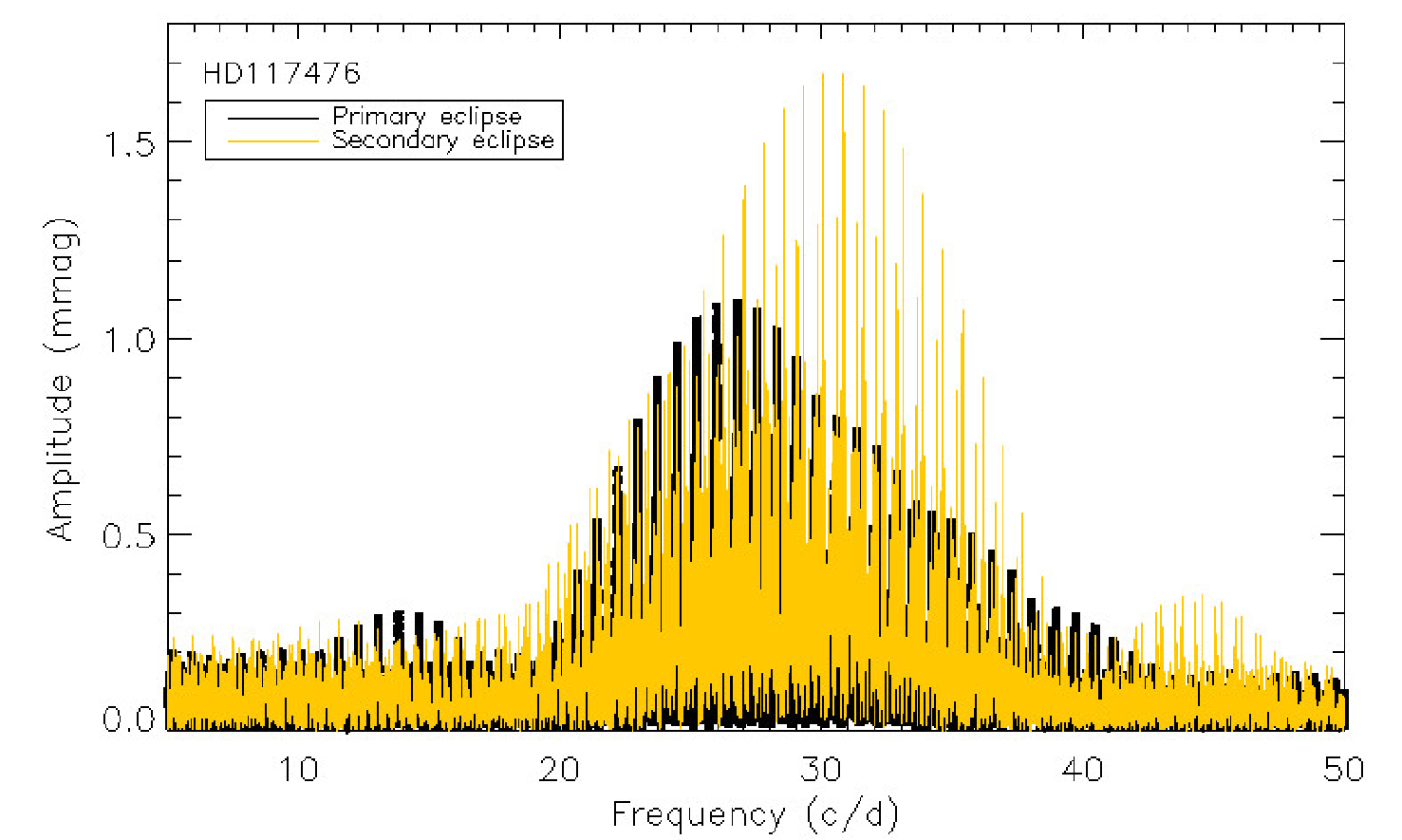}
 \end{minipage}
 \begin{minipage}[b]{0.45\textwidth}
  \includegraphics[height=4.5cm, width=1\textwidth]{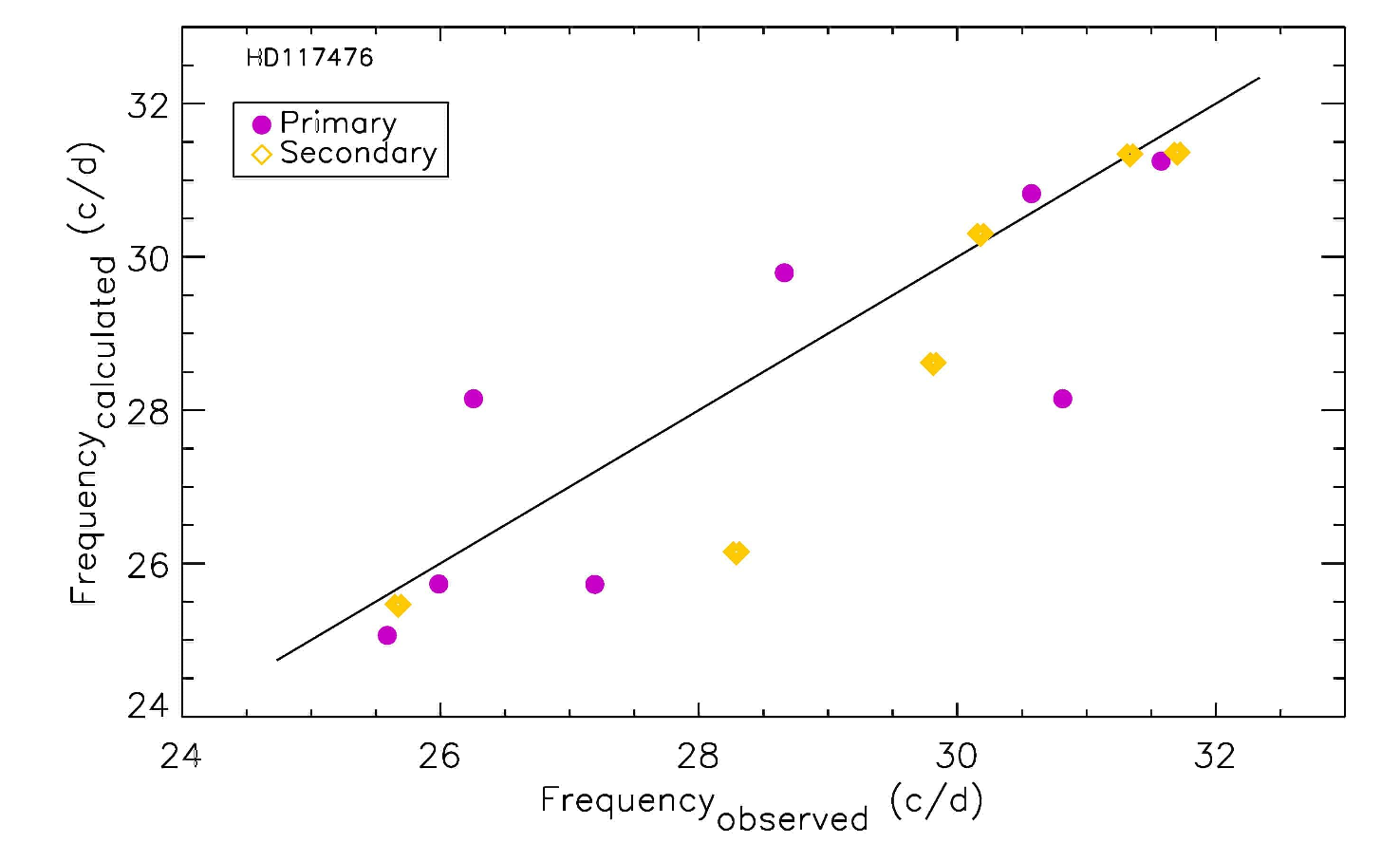}
  \end{minipage}
   \begin{minipage}[b]{0.45\textwidth}
  \includegraphics[height=4.5cm, width=1\textwidth]{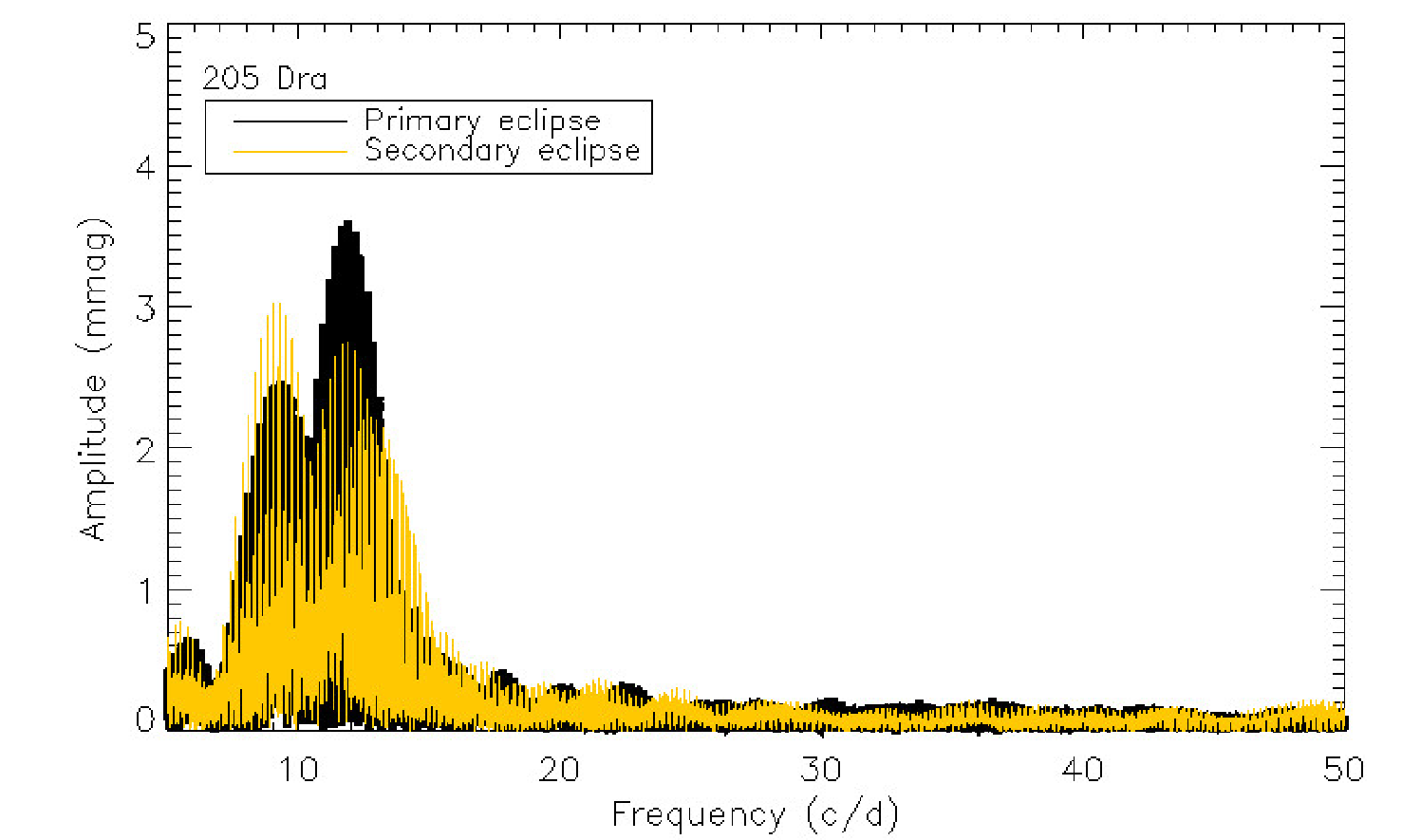}
  \end{minipage}
   \begin{minipage}[b]{0.45\textwidth}
  \includegraphics[height=4.5cm, width=1\textwidth]{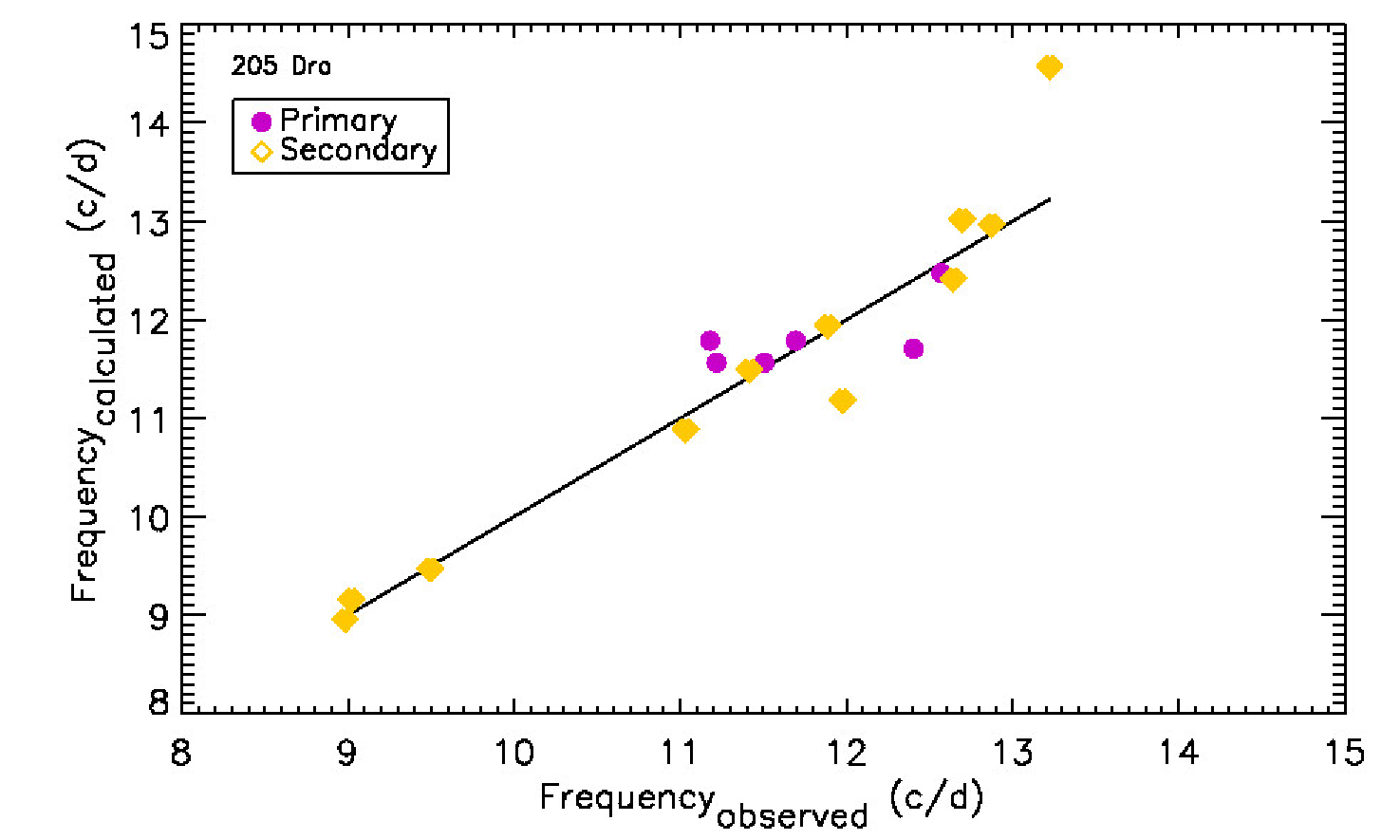}
  \end{minipage}
    \begin{minipage}[b]{0.45\textwidth}
  \includegraphics[height=4.5cm, width=1\textwidth]{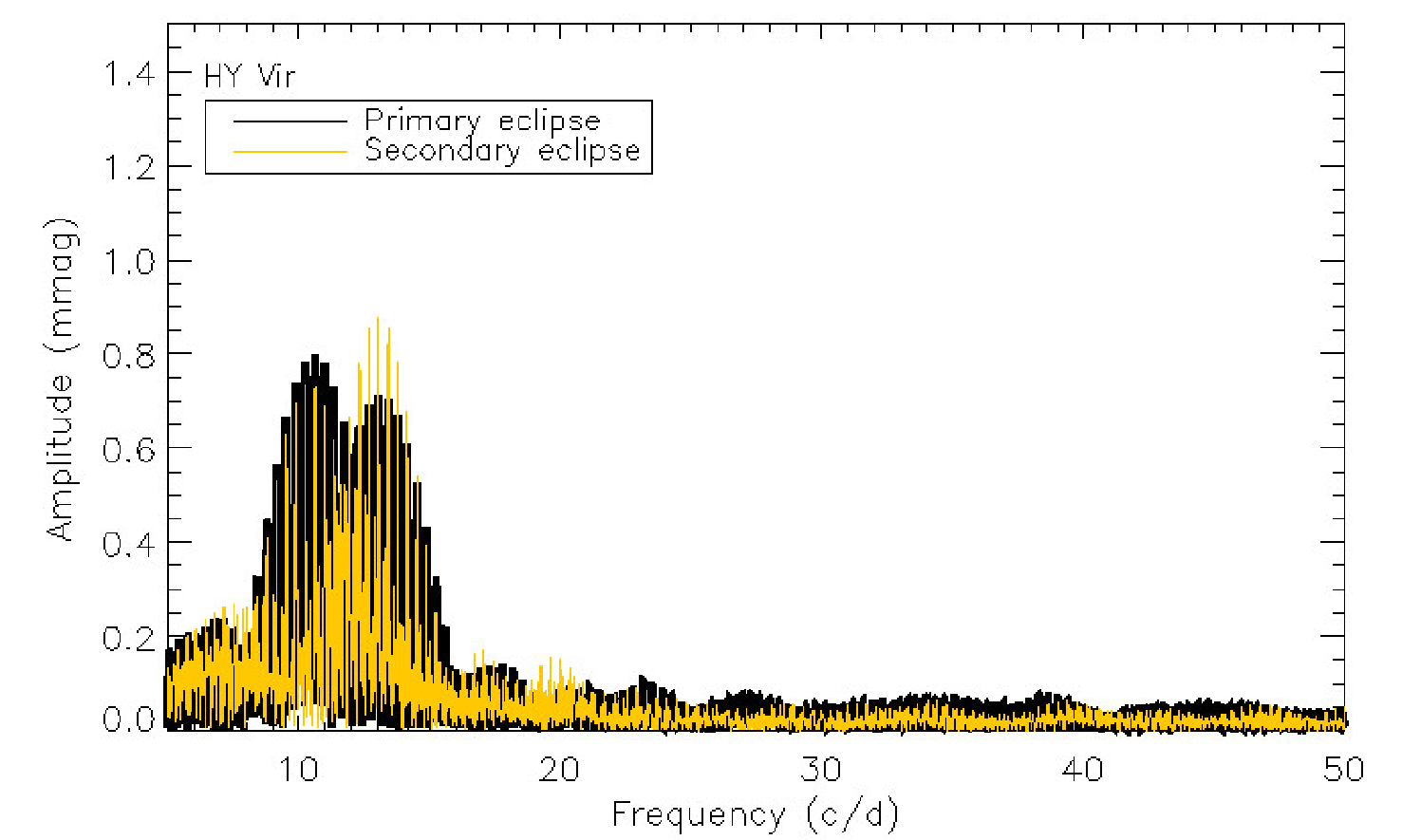}
  \end{minipage}
    \begin{minipage}[b]{0.45\textwidth}
  \includegraphics[height=4.5cm, width=1\textwidth]{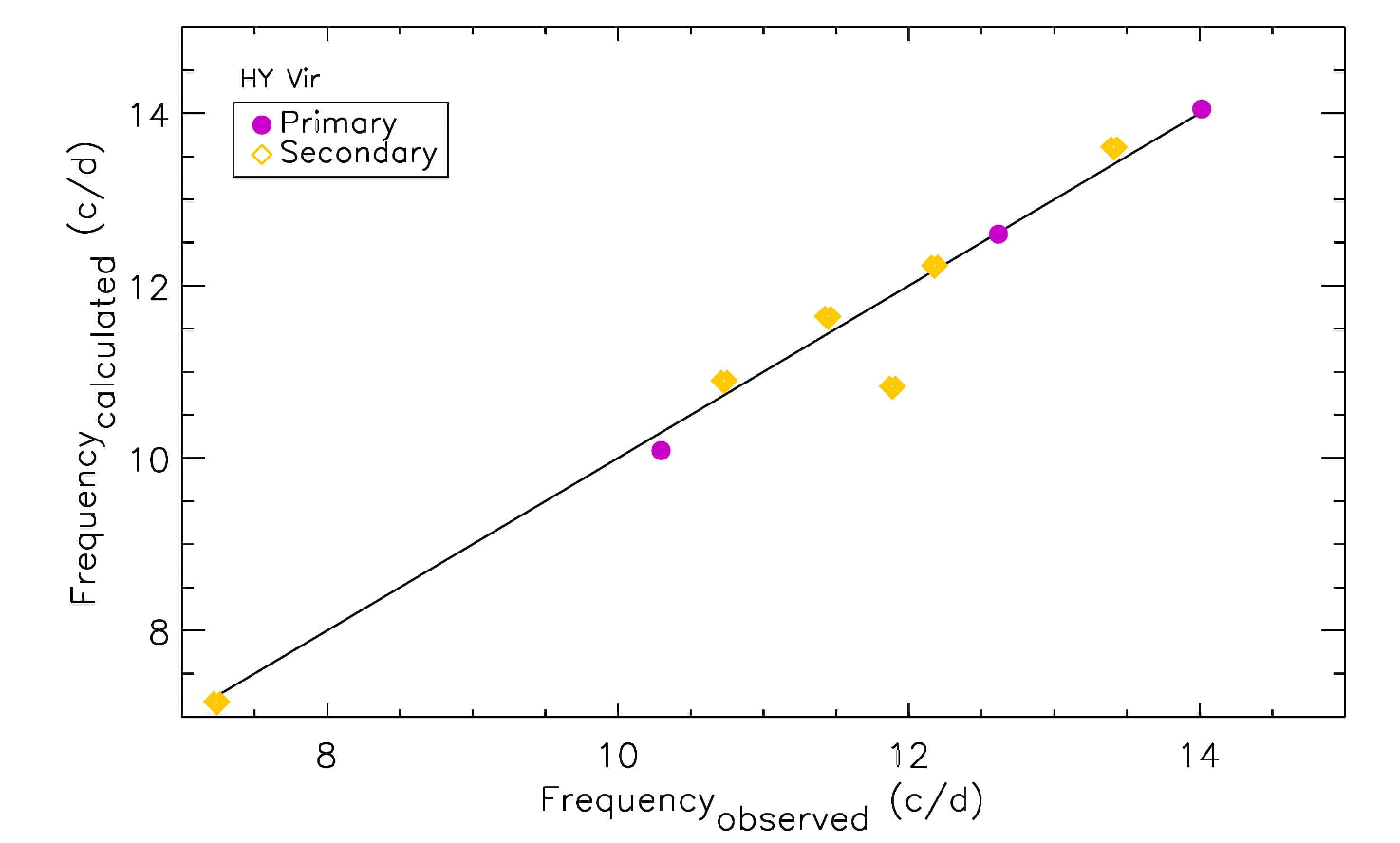}
  \end{minipage}
      \begin{minipage}[b]{0.45\textwidth}
  \includegraphics[height=4.5cm, width=1\textwidth]{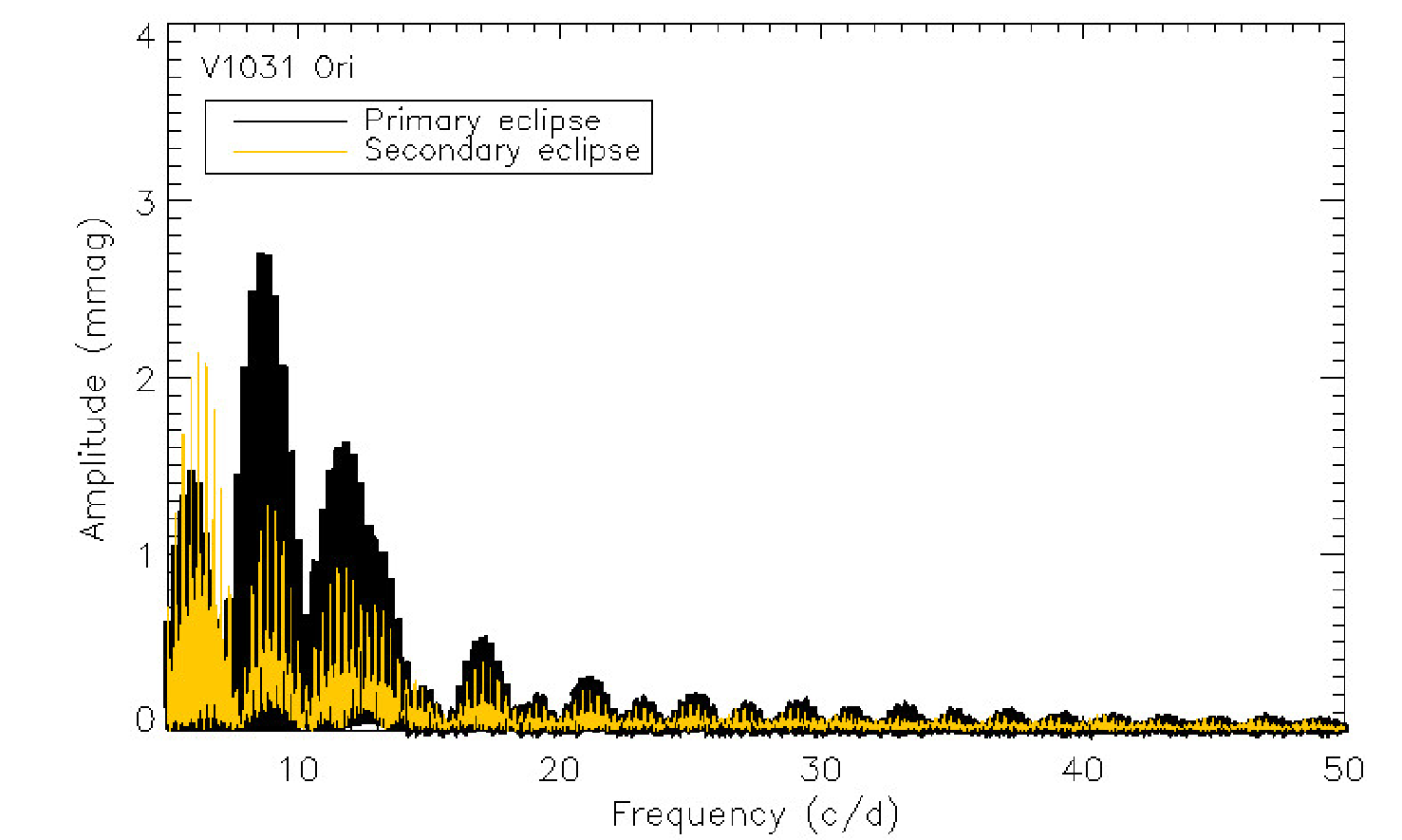}
  \end{minipage}
    \begin{minipage}[b]{0.45\textwidth}
  \includegraphics[height=4.5cm, width=1\textwidth]{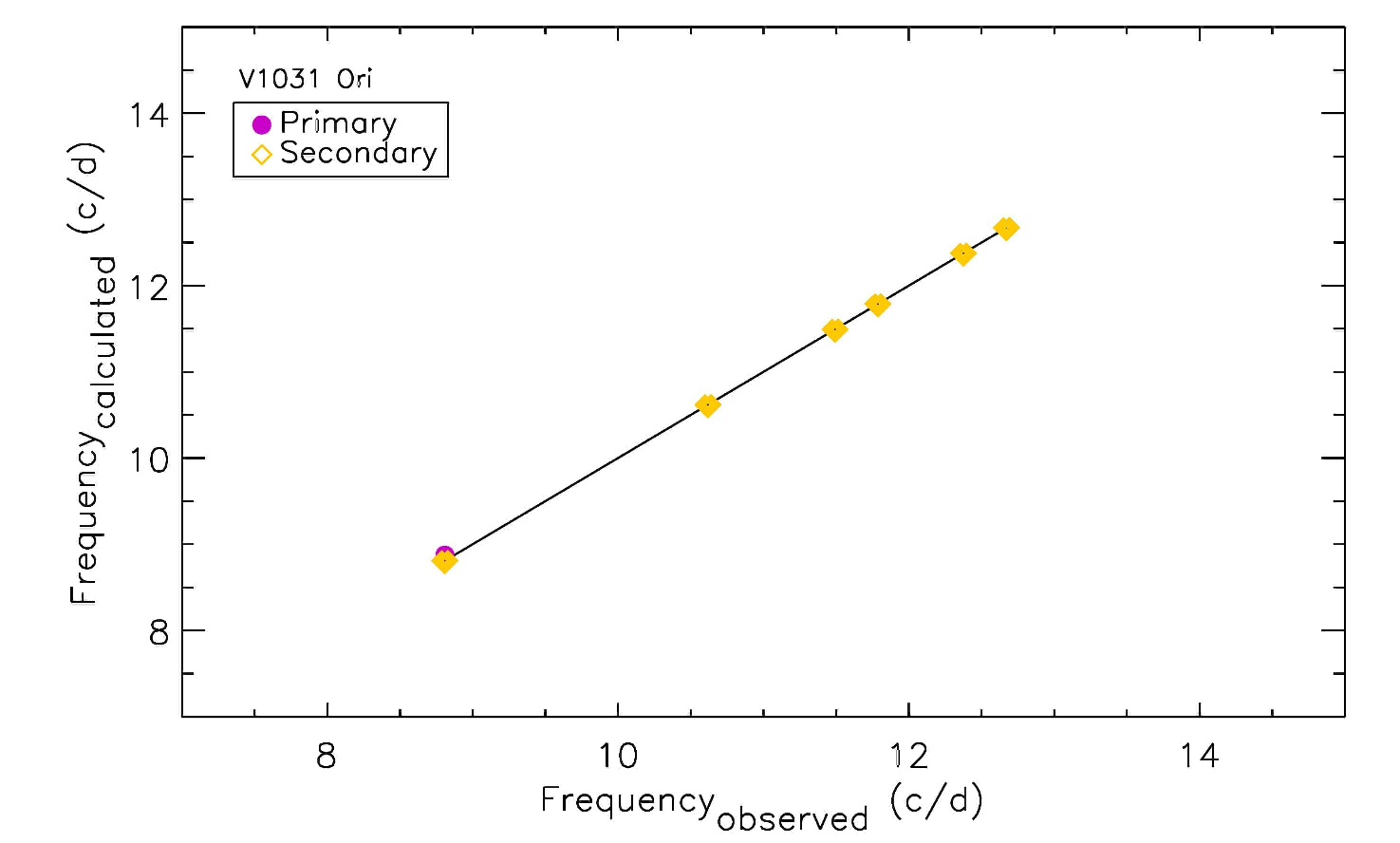}
  \end{minipage}
   \caption{Left panels: Amplitude spectra derived from the data obtained during the primary and secondary eclipses for each target. Right panels: Comparison of the theoretical frequencies calculated using GYRE models with the observed frequencies. The labels "primary" and "secondary" in these panels have the same meaning as in Fig.\,3.}  \label{fig:pulscont}
\end{figure*}

In the second approach, the frequency amplitudes of the targets during the primary and secondary eclipses were calculated and compared to estimate which component of the binary is pulsating. During the primary (deeper) eclipse, the primary component is obscured by the secondary, and the observed light predominantly comes from the secondary star. In contrast, during the secondary eclipse, the light is mainly contributed by the primary component. By analyzing the pulsation amplitudes around the times of the primary and secondary eclipses, it is possible to infer which component is pulsating based on the variations in amplitude observed during these phases. In the case of V1031\,Ori, we were unfortunately unable to distinguish the contribution of the third light. However, based on our examination in the following and previous approaches, we concluded that the third component is probably not the pulsating one. Therefore, in the current analysis, we applied the method to V1031\,Ori assuming that one of the other binary components is the likely pulsator. In this approach, we first noticed that all systems exhibit similar pulsation spectra during each eclipse, though with varying amplitudes. This suggests that most likely only one component is responsible for the pulsations in these binary systems. The pulsation amplitude spectra for each target at primary and secondary eclipses are shown in Fig.\,\ref{fig:pulscont}. According to the amplitude spectra, for HD\,117476, the pulsation amplitude at secondary eclipse is significantly higher than primary eclipse. Therefore, in this second approach, we could say that the primary component is likely the pulsating component of HD\,117476. In the case of 205\,Dra, the pulsation amplitude is higher at primary eclipse, hence, the secondary component seems to the pulsating component in this system. For HY\,Vir, the amplitude of pulsations at primary eclipse are higher comparing to secondary eclipse. Therefore, the secondary component in HY\,Vir is accepted as pulsator with the approach. For the last system V1031\,Ori, the pulsation amplitude at primary eclipse seem generally higher comparing to secondary one, hence, the secondary component is assumed to be pulsator in this system.

In addition to these approaches, we also planed to calculate theoretical frequencies which appear from the binary components of each target and compare them with the observation to find out the which components pulsate. We performed non-adiabatic pulsation analyses for our targets using the \texttt{GYRE} (version 7.2.1) pulsation code \citep{2013MNRAS.435.3406T, 2018MNRAS.475..879T}. A linear grid scan was carried out over the 10–1000\,$\mu$Hz range with parameters $w_{osc}$ = 10, $w_{exp}$ = 2, and $w_{ctr}$ = 10. The input metallicities and fundamental parameters were adopted from Tables\,\ref{tab:atmospar} and \ref{tab:lc}. For each target, we selected the best-fitting models that most accurately reproduce the observed non-asteroseismic parameters and used them as input for the \texttt{GYRE} calculations.

We computed non-adiabatic oscillation frequencies to identify excited modes and assess their stability. Initial trial frequencies were generated using various built-in methods in \texttt{GYRE}, allowing us to analyze non-radial oscillation modes. A comparison between the model and observed frequencies shows good agreement, confirming the reliability of our models and supporting the validity of the adopted approach for $\delta$ Scuti-type pulsators.

The consistency between the theoretical and observed frequencies for each target is illustrated in Fig.\,\ref{fig:pulscont}. Based on this comparison, we determined the pulsating component in each system, as listed in Table\,\ref{tab:puls_comp} alongside results from other approaches. If at least two approaches yielded consistent results, we selected that component as the pulsating one. Additionally, the \texttt{GYRE} models did not predict any frequencies for the third component of V1031\,Ori that could explain the observed pulsations. Therefore, we consider the third component of V1031\,Ori to be non-pulsating star. As can be seen from Table\,\ref{tab:puls_comp}, the second and third approaches are mostly consistent with each other, while the first approach differs. This is likely because the first method, based on the pulsation constant, is a simplified approximation that does not account for detailed stellar structure, rotation, or non-radial modes.

\begin{table}
\centering
  \caption{Comparison of the used approaches for estimating the pulsating binary components in the targets.}
  \label{tab:puls_comp}
  \begin{tabular}{lccc}
\hline
Target             & Approach 1   &  Approach 2     &  Approach 3   \\
\hline
HD117476   &-   & primary    & primary\\
205\,Dra   &primary   & secondary    & secondary\\
HY\,Vir    &secondary   & secondary    & secondary\\
V1031\,Ori &primary   & secondary    & secondary\\
\hline
\end{tabular}
\end{table}

\section{Evolutionary modeling}\label{sect:evo}

\begin{figure*}
\centering
\begin{minipage}[b]{0.45\textwidth}
 \includegraphics[height=6cm, width=1\textwidth]{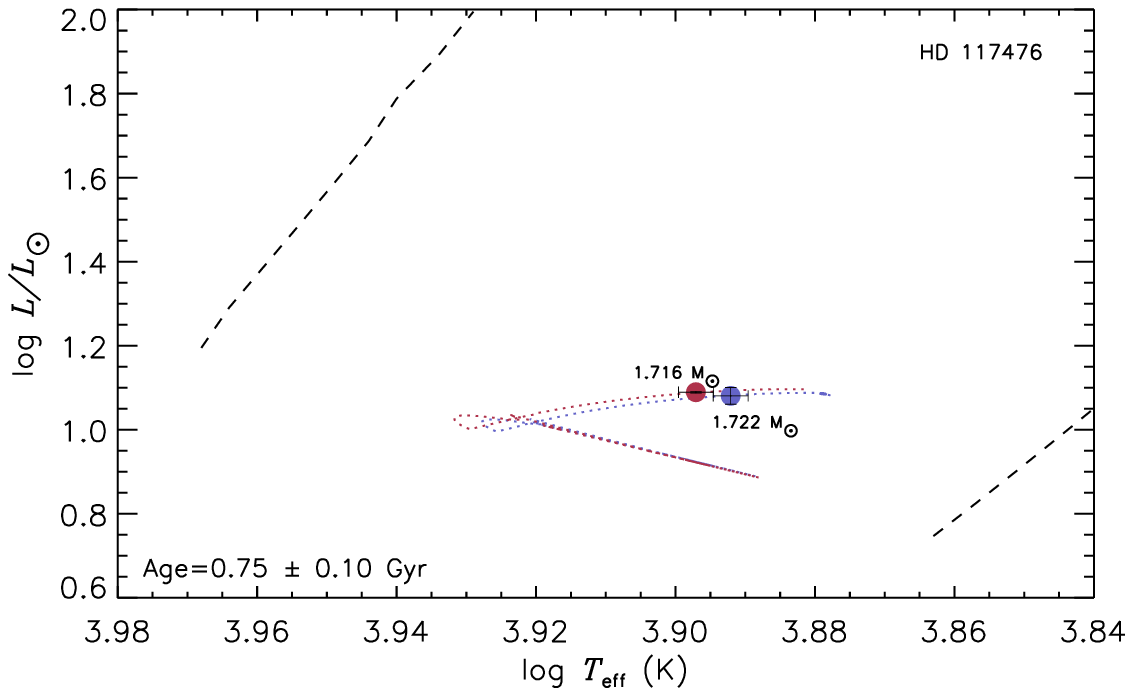}
\end{minipage}
\begin{minipage}[b]{0.45\textwidth}
\includegraphics[height=6cm, width=1\textwidth]{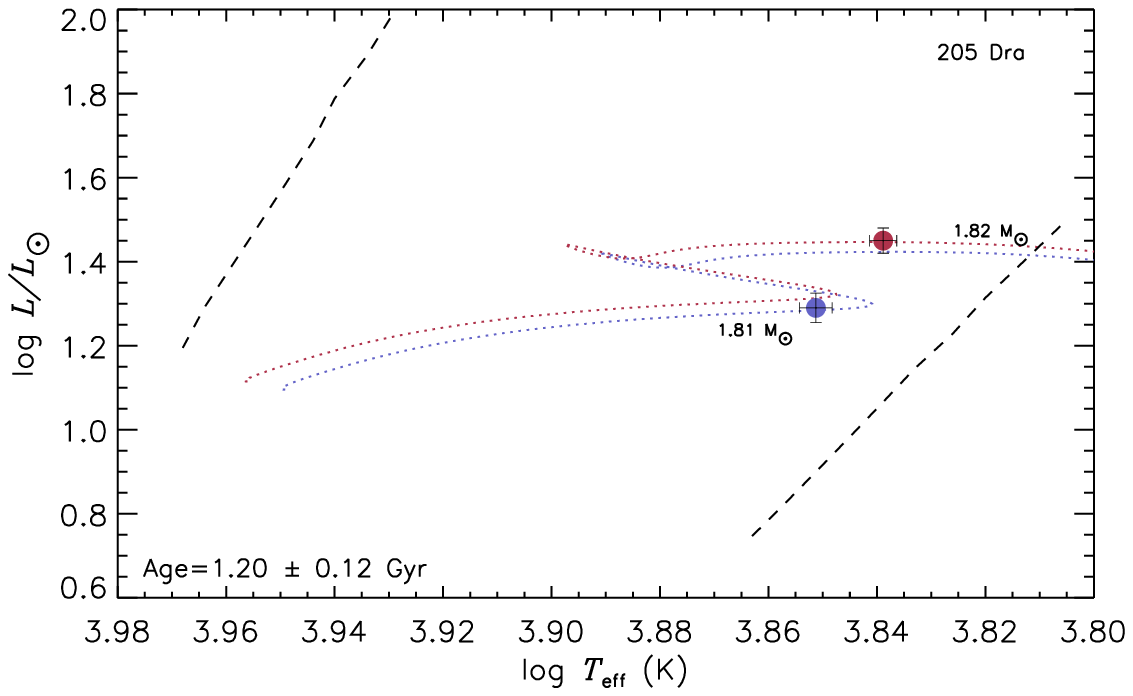}
\end{minipage}
 \begin{minipage}[b]{0.45\textwidth}
\includegraphics[height=6.5cm, width=1\textwidth]{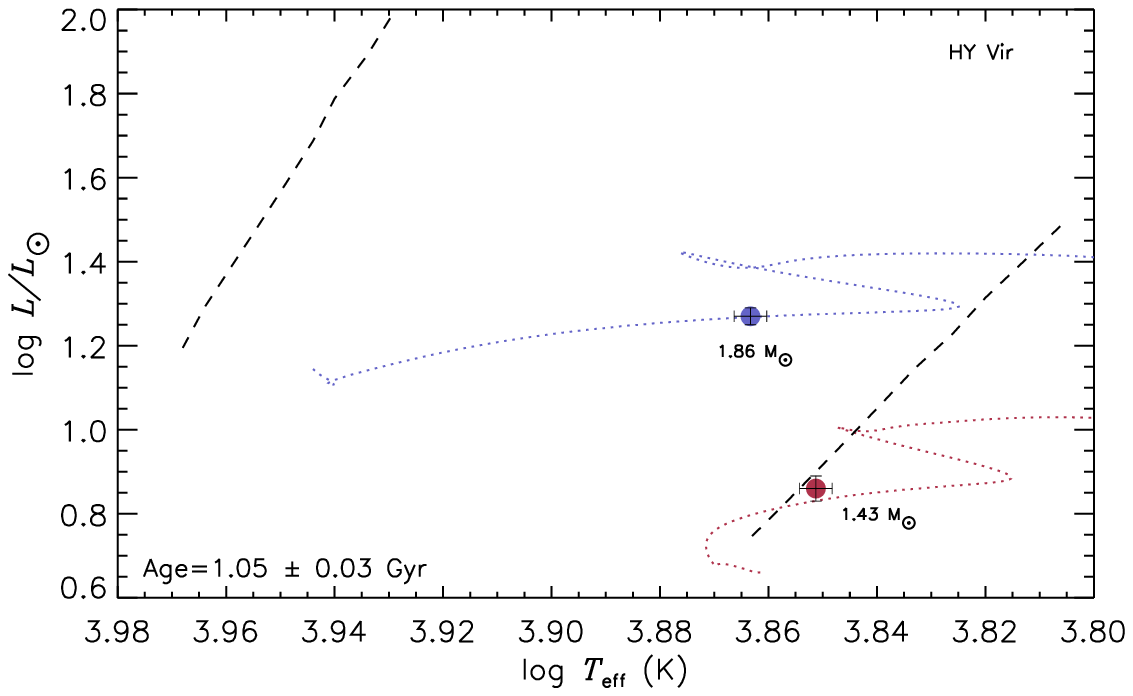}
\end{minipage}
 \begin{minipage}[b]{0.45\textwidth}
\includegraphics[height=6.5cm, width=1\textwidth]{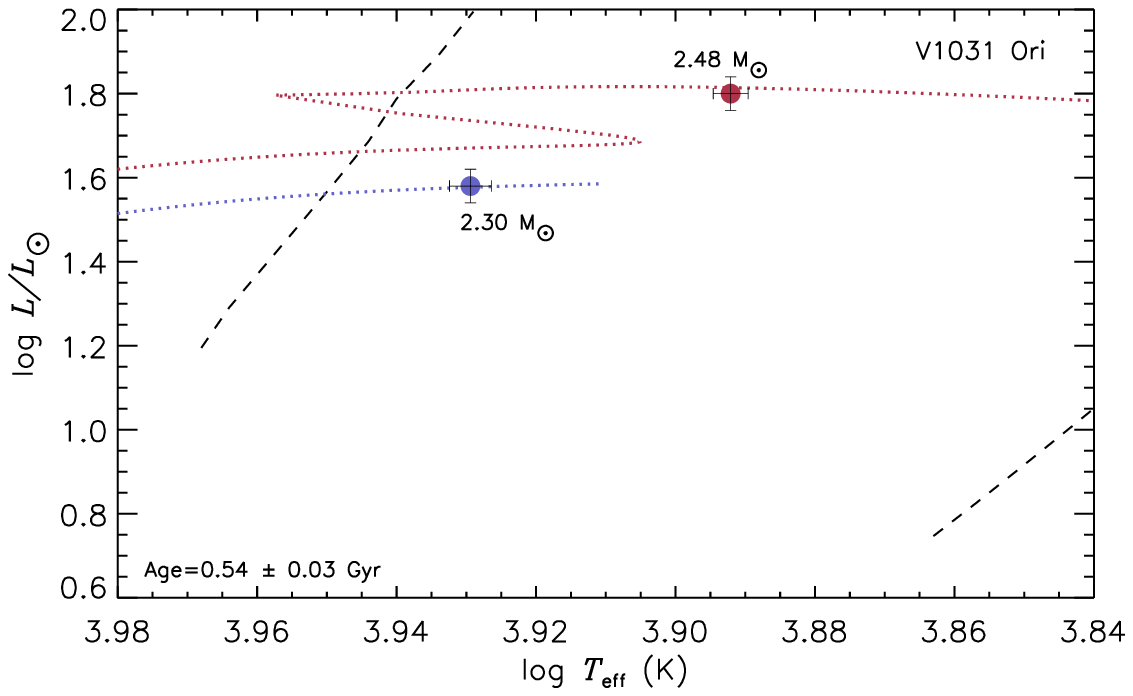}
\end{minipage}
\caption{Positions of the targets on the H-R diagram. Blue and red dots represent the primary and secondary components, respectively. The dotted and dashed lines are the MESA binary evolutionary tracks and the observational borders of the \ds\, instability strip \citep{2019MNRAS.485.2380M}, respectively.}  \label{fig:hr}
\end{figure*}

The evolutionary status of the targets was investigated using the
{\small {binary}} module in the Modules for Experiments in Stellar Astrophysics {\small {MESA}} code \citep{2011ApJS..192....3P, 2013ApJS..208....4P}.
 The binary module \citep{2015ApJS..220...15P} analyze the binary orbital evolution and determine the initial parameters of binary systems. In this examination, MESA equation-of-state (EOS) based on the OPAL EOS tables was used \citep{2002ApJ...576.1064R}. During the analysis, the initial metallicity value ($Z$) was adopted as solar \cite{2009ARA&A..47..481A} and helium mass fraction was taken 0.28 for this $Z$ value. The convective core overshoot and the mixing length, $\alpha_{MLT}$, value were taken as 0.20 and 1.8 in the investigation considering the values used for the modeling of the borders of the theoretical \ds\, instability strip \citep{2005A&A...435..927D}. 
Various evolutionary models were obtained with different initial $M$, $P_{orb}$, rotational velocity ($v_{rot}$), and $Z$ values until finding the best-fitting model with the current evolutionary status of the systems. The resulting parameters of the evolutionary modeling is given in Table\,\ref{tab:evoltable}. During this analysis the $Z$ value was not assumed to be the same for both components of a binary system. Since our spectral analysis revealed different metallicity values for the binary components of the same system, we aimed to verify this result through the current evolutionary analysis. As can be seen from Table\,\ref{tab:evoltable}, the results are mostly consistent with the spectral analysis. The discrepancy between the spectroscopically derived Fe abundances and the $Z$ estimated in the evolutionary models can be attributed to the fact that $Z$ reflects the overall metal content of the star, while the Fe abundance is a surface property that can be altered by atomic diffusion and radiative levitation. Especially in chemically peculiar stars, such as Am stars, surface Fe enhancement does not necessarily imply a higher global metallicity \citep{2019MNRAS.485.1067T, 2005MNRAS.363..529S} The best fitting evolutionary models with the observation for each targets are shown in Fig.\,\ref{fig:hr}. The ages of the systems were also estimated by looking at model age at the position of the targets on the H-R diagram. 




\begin{table*}
\centering
\caption{Results obtained from the best-fit evolutionary models.}
\begin{tabular*}{0.95\linewidth}{@{\extracolsep{\fill}}lcccc}
\hline
  Parameter                & HD\,117476   & 205\,Dra    & HY\,Vir  & V1031\,Ori\\
                           &              &             &          & \\
\hline
$M_p$$_{initial}$  ($M_\odot$)    & 1.722 (3)     & 1.81 (5) & 1.86 (2) & 2.30 (2) \\
$M_s$$_{initial}$  ($M_\odot$)    & 1.716 (3)     & 1.82 (5) & 1.43 (2) & 2.48 (2) \\
$P_{orb}$$_{initial}$  (days)    & 2.14 (3)     & 4.47 (5)    & 3.22 (2) & 3.69 (2)\\
$Z$$_{p}$                  & 0.014 (2)    & 0.012 (2)   & 0.014 (2) & 0.020 (2)\\
$Z$$_{s}$                  & 0.013 (2)    & 0.013 (2)   & 0.014 (2) & 0.017 (2)\\
Age (Gyr)                  &0.75 (8)     & 1.20 (12)    & 1.05 (6)  & 0.54 (3) \\
\hline
\end{tabular*}
\label{tab:evoltable}
\end{table*}

\section{Discussion and Conclusions}

In this study, we focused on four detached eclipsing binaries, HD117476, 205\,Dra, HY\,Vir, V1031\,Ori, that exhibit $\delta$ Scuti-type oscillations. These targets were selected because all their components were found to be located within the $\delta$ Scuti instability strip. Our aim was to identify which components in each system are pulsating and which are not, and to investigate the differences between the pulsating and non-pulsating components within the same binary system. For these purpose, we carried out detailed photometric and spectroscopic analysis using the TESS data and high-resolution spectra. 

The orbital and atmospheric parameters of all targets were derived in this study. For HD\,117476 and 205\,Dra, such a detailed spectroscopic analysis was performed for the first time. HY\,Vir and V1031\,Ori had been examined previously in the literature. For HY\,Vir, \citet{2008NewA...13..304M} estimated the \teff\, values based on spectral type and photometry, adopting 7095\,K for the primary component. Our spectroscopic analysis yielded \teff\, values that agree with theirs within the error bars. Additionally, \citet{2011AJ....142..185S} measured and analyzed the radial velocity (RV) variations of HY\,Vir. We included their measurements in our study and found our RV analysis to be consistent with their results. V1031\,Ori was previously investigated by \citet{1990A&A...228..365A}, whose RV data we also adopted. Since we only used three spectra for V1031\,Ori, our RV analysis produced similar parameters within the uncertainties.

After atmospheric parameter determination we obtained chemical abundances of the binary components. We carried out this analysis only for the stars (HD\,117476, 205\,Dra, HY\,Vir) whose individual binary component spectra were obtained from spectral disentangling. As a result of this analysis, we found that the some binary component show higher Fe abundance comparing the other component in the same binary system. This result is unexpected for the binary system in which the binary components exist from the same interstellar environment. Recent analyses using disentangled spectra, such as those by \citet{2019MNRAS.485.1067T}, emphasized that chemical peculiarities can coexist with pulsations, especially in slow rotators within binary systems. There are also other studies which show that two binary components could have different chemical composition \citep[e.g.][]{2015AJ....149...59Y, 2020RAA....20..150K, 2023MNRAS.520.1601K}. This situation could be explained that the accretion of gas and dust from the surrounding circumstellar envelope might modify the atmospheric composition of one of the components \citep{2015AJ....149...59Y}.

The binary modeling of the systems was performed following the analysis of radial velocities and atmospheric parameters. Fundamental astrophysical parameters were derived for all targets. For HY\,Vir \citep{2011AJ....142..185S} and V1031\,Ori \citep{1990A&A...228..365A, 2021PASJ...73..809L}, these parameters are also available in the literature. In the case of V1031\,Ori, the parameters estimated by \citet{2021PASJ...73..809L} are consistent with our results within the uncertainties, except for the $R$ of the secondary component. In both studies, TESS data were used; however, the difference arises from the treatment of pulsation effects and the adopted \teff\, values. In our analysis, we removed the pulsation signals prior to the binary modeling, which explains the discrepancy and is considered acceptable. For HY\,Vir, the fundamental parameters differ significantly from those reported by \citet{2011AJ....142..185S}, likely due to their use of ground-based photometric data. We also repeated the abundance analysis using the more precise \logg\, values from the binary modeling (Sect. 4.1, Table 5). The absolute Fe abundances changed by about 0.02–0.03\,dex for most components, and the largest change was 0.09\,dex for the secondary of 205 Dra. These small differences are within the uncertainties of the atmospheric analysis and do not change our main results.

The evolutionary status of the systems was also examined. Through this analysis, we estimated the ages of the systems, as well as their initial $M$ and initial orbital $P_{\rm orb}$. According to these models, the systems are plotted on the H-R diagram, and, as previously estimated, all binary components were found to lie within the \ds\, instability strip.

We investigated the pulsation behavior of the binary components. Oscillation frequencies were obtained for each system. Since all components of the selected binaries lie within the $\delta$ Scuti instability strip, we carried out a series of analyses to determine which components are pulsating. Three different approaches were applied, and in each case, only one component in each system was identified as the pulsator. We found that in HD\,117476, the primary component is pulsating, while in the other targets, the secondary components were found to be the pulsators. We investigated the differences between the pulsating and non-pulsating components within the same binary systems. It turned out that the pulsating components are chemically normal stars, while the non-pulsating ones are chemically peculiar stars in the cases of HD\,117476, 205\,Dra, and HY\,Vir. Although our study shows that stars exhibiting Am characteristics do not pulsate, it is important to note that pulsating $\delta$ Scuti stars with Am-like abundance patterns have been reported in the literature (e.g., \citealt{2005MNRAS.359..865Y}; \citealt{2008CoSka..38..123F}; \citealt{2003A&A...398.1121E}). However, most of these previous studies focused on single stars or wide binaries where tidal interactions and common formation environments are not present. In contrast, our targets are close, eclipsing binary systems with significant dynamical interaction, and their components are expected to have formed under the same physical conditions. Therefore, the presence of non-pulsating Am components and pulsating chemically normal companions within the same system offers a valuable opportunity to explore the impact of chemical peculiarity on pulsation in a controlled environment. Such systems may help to better understand whether and how Am characteristics suppress pulsations under binary evolutionary effects, a topic also addressed in spectroscopic investigations by \citet{2019MNRAS.485.1067T} and others.

Since we could not estimate the metallicity of V1031\,Ori from our atmospheric models, we attempted to determine whether there is a $Z$ difference in this system through our evolutionary modeling. As a result, we found the same pattern: the pulsating component appears chemically normal, while the other (primary) component shows a higher $Z$ value. Previous investigations have suggested that, theoretically, chemically peculiar (metal-rich) A-type stars should not exhibit \ds\,-type oscillations \citep{1976ApJS...32..651K, 2000A&A...360..603T}. The main reason for this is that chemical separation (enrichment or depletion of elements through diffusion) suppresses the convective zones in the stellar interior. These convective regions are essential for driving mechanisms such as the $\kappa$ mechanism, which powers the oscillations. In other words, metal enrichment alters the opacity and weakens convection in the energy transport process, which is expected to suppress pulsations. However, there are studies that do not fully support this theory \citep[e.g.][]{2019MNRAS.484.2530C, 2024A&A...685A.133C, 2024A&A...690A.104D, 2024MNRAS.534.3211S}.

Our targets were selected based solely on the positions of their components within the \ds\, instability strip, meaning there was no selection bias. While our investigation shows that the pulsating components are chemically normal stars and the non-pulsating components are metal-rich, there are studies in the literature reporting systems where both components are metal-rich yet still exhibit oscillations \citep[e.g.][]{2019MNRAS.487..919C, 2024A&A...685A.133C}. \citet{2000A&A...360..603T} suggested that pulsations are absent in young Am stars but tend to develop naturally as these stars evolve away from the Zero-Age Main Sequence (ZAMS). Such evolutionary differences might explain the discrepancy between our findings and those of other studies.

These investigations are crucial for understanding the nature of oscillating stars and the influence of chemical composition on pulsations. Since our targets are detached eclipsing binaries, they can be treated similarly to single stars (as binary effects are negligible in these systems), and they allow us to derive precise fundamental parameters. Therefore, examining \ds\,-type oscillations in such detached binary systems will significantly contribute to a better understanding of the role of metallicity in the $\kappa$ mechanism.

As a future plan, we intend to continue analyzing such systems to further investigate the impact of metallicity, by comparing our results with those reported in the literature.

\section*{Acknowledgements}

We thank the referee for the constructive comments and helpful suggestions that improved the paper. This study has been supported by the Scientific and Technological Research  Council (TUBITAK) project through 120F330. The calculations have been carried out in Wroc{\l}aw Centre for Networking and Supercomputing (http://www.wcss.pl), grant No.\,214. Based on data obtained from the ESO Science Archive Facility. The TESS data presented in this paper were obtained from the Mikulski Archive for Space Telescopes (MAST). Funding for the TESS mission is provided by the NASA Explorer Program. This work has made use of data from the European Space Agency (ESA) mission Gaia (http://www.cosmos.esa.int/gaia), processed by the Gaia Data Processing and Analysis Consortium (DPAC, http://www.cosmos.esa.int/web/gaia/dpac/consortium). Funding for the DPAC has been provided by national institutions, in particular, the institutions participating in the Gaia Multilateral Agreement. This research has made use of the SIMBAD data base, operated at CDS, Strasbourq, France.

\newpage

\appendix

\setcounter{table}{0}
\renewcommand{\thetable}{A\arabic{table}}

\begin{table*}
\begin{center}
\centering 
\caption[]{The RV measurements. The subscripts ``p'' and ``s'' represent the primary and the secondary components, respectively.}\label{tab:A1}

\begin{tabular*}{0.95\linewidth}{@{\extracolsep{\fill}}lcc|ccc}
\hline
   & HD\,117476 &               &      &205\,Dra   &            \\
   &            &               &      &             &             \\
\hline
  HJD (2450000+)         & RV$_{p}$ & RV${_s}$ & HJD (2450000+) &RV$_{p}$  & RV${_s}$ \\
\hline             
9691.4658    &-49.9\,(2)   &              &9731.5854  &-12.1\,(1) &-12.1\,(1)  \\
9692.4582    &-138.8\,(2)  &107.4\,(3)    &9732.5744  &-106.1\,(3)&86.2\,(4)  \\ 
9695.4458    &-23.3\,(2)   &              &9747.5467  & 92.3\,(3) &-105.8\,(3)  \\
9714.3981    & 68.2\,(1)   &-98.7\,(2)    &9755.5016  &60.7\,(3) &-82.1\,(3) \\
9715.3853    &91.5\,(2)    &-128.1\,(3)   &9756.4906  &68.0\,(2) &-86.4\,(3) \\
9732.3436    & 22.0\,(2)   &              &9757.5342  &-69.3\,(2) &54.1\,(3) \\
9733.3396    &-148.1\,(2)  &101.9\,(3)    &9765.4790  &-11.7\,(1) &-11.7\,(1)  \\
9747.4029    & -23.6\,(1)  &              &9766.5422  & -108.0\,(3) &86.7\,(3) \\
9766.3661    & -60.7\,(3)  &              &  & &            \\
9769.3594    &108.7\,(1)   &-152.3\,(3)   &  & &            \\
 \hline
 \hline
       &            &               &      &             &             \\
   & HY\,Vir &               &      &V1031\,Ori   &            \\
   &            &               &      &             &             \\
\hline
  HJD (2450000+)         & RV$_{p}$ & RV${_s}$ & HJD (2450000+)           &RV$_{p}$  & RV${_s}$ \\
\hline             
4887.8012    &61.2\,(2)   &  -113.6\,(3)  &6909.8101  &112.2\,(1) &-112.2\,(1)  \\
4889.7759    &-85.0\,(2)  &90.6\,(3)      &6910.8208  &-70.6\,(1)&67.7\,(1)  \\ 
4889.8550    &-71.82\,(2) &75.3\,(2)      &6910.9148  &-88.6\,(1) &83.1\,(2)  \\
4890.7551    & 79.8\,(2)   &-132.5\,(4)   &  & &            \\
4890.7958    &82.5\,(1)    &-131.7\,(3)   &  & &            \\
4890.8556    &81.4\,(2)    &-132.4\,(2)   &  & &            \\
6449.5210    &-107.2\,(2)  &105.8\,(4)    &  & &            \\
7006.8801    &-103.7\,(1)  &110.3\,(4)    &  & &            \\
7008.8801    & 30.3\,(3)  & -82.1\,(3)    &  & &            \\
7029.8690    & 48.8\,(3)   &-78.3\,(4)  &  & &            \\
\hline
\end{tabular*}
     \end{center}
\end{table*}

\begin{table*}
\centering
\label{abundancetable}
\caption{Abundances of individual elements of the binary components of HD117476, 205\,Dra, and HY\,Vir. The subscripts ``p'' and ``s'' represent the primary and the secondary components, respectively. Number of analyzed parts is given in brackets.}\label{tab:A2}
\begin{tabular*}{0.9\linewidth}{@{\extracolsep{\fill}}lllllll}
\hline
  Elements& HD\,117476$_{p}$       & HD\,117476$_{s}$     & 205\,Dra$_{p}$       &   205\,Dra$_{s}$  &   HY\,Vir$_{p}$ &   HY\,Vir$_{s}$\\
\hline   
$_{12}$Mg & 7.85\,$\pm$\,0.43\,(5) & 8.33\,$\pm$\,0.54\,(3) &  8.31\,$\pm$\,0.50\,(2) & 7.26\,$\pm$\,0.42\,(3) & 8.44\,$\pm$\,0.51\,(2) & 7.77\,$\pm$\,0.50\,(3)\\
$_{14}$Si & 7.62\,$\pm$\,0.42\,(4) & 7.36\,$\pm$\,0.44\,(5) &  7.90\,$\pm$\,0.74\,(8) & 7.19\,$\pm$\,0.67\,(8) & 8.54\,$\pm$\,0.42\,(3) & 8.64\,$\pm$\,0.51\,(2)\\
$_{20}$Ca & 6.37\,$\pm$\,0.48\,(11)& 5.96\,$\pm$\,0.48\,(3) &  6.28\,$\pm$\,0.89\,(3) & 5.80\,$\pm$\,0.38\,(6) & 6.73\,$\pm$\,0.39\,(5) & 5.90\,$\pm$\,0.42\,(8)\\
$_{21}$Sc & 2.05\,$\pm$\,0.47\,(3) & 3.08\,$\pm$\,0.46\,(2) &  3.36\,$\pm$\,0.72\,(2) & 2.40\,$\pm$\,0.43\,(2) & 3.40\,$\pm$\,0.38\,(4) & 3.55\,$\pm$\,0.58\,(5)\\
$_{22}$Ti & 5.03\,$\pm$\,0.41\,(16)& 5.81\,$\pm$\,0.41\,(12)&  5.41\,$\pm$\,0.48\,(13)& 5.03\,$\pm$\,0.52\,(24)& 5.70\,$\pm$\,0.39\,(28)& 5.14\,$\pm$\,0.38\,(39)\\
$_{24}$Cr &5.92\,$\pm$\,0.44\,(14) & 6.60\,$\pm$\,0.49\,(6) &  6.53\,$\pm$\,0.82\,(24)& 6.03\,$\pm$\,0.47\,(20)& 6.66\,$\pm$\,0.42\,(30)& 6.03\,$\pm$\,0.40\,(30)\\
$_{25}$Mn &5.46\,$\pm$\,0.40\,(5)  & 5.88\,$\pm$\,0.48\,(2) &  6.22\,$\pm$\,0.56\,(3) & 4.79\,$\pm$\,0.53\,(2) & 6.04\,$\pm$\,0.41\,(7) & 5.98\,$\pm$\,0.42\,(9)\\
$_{26}$Fe &7.38\,$\pm$\,0.32\,(51) & 8.22\,$\pm$\,0.34\,(41)&  8.24\,$\pm$\,0.32\,(75)& 7.50\,$\pm$\,0.35\,(92)& 8.17\,$\pm$\,0.33\,(68)& 7.58\,$\pm$\,0.37\,(79)\\
$_{28}$Ni &6.46\,$\pm$\,0.44\,(14) & 7.05\,$\pm$\,0.49\,(9) &  7.26\,$\pm$\,0.46\,(16)& 6.56\,$\pm$\,0.47\,(20)& 7.35\,$\pm$\,0.39\,(27)& 6.47\,$\pm$\,0.40\,(28)\\
$_{29}$Cu &                        &                        &  4.52\,$\pm$\,0.38\,(1) & 3.87\,$\pm$\,0.45\,(1) & 5.97\,$\pm$\,0.50\,(1) & 3.61\,$\pm$\,0.47\,(2)\\
$_{30}$Zn &                        & 5.42\,$\pm$\,0.51\,(5) & 5.27\,$\pm$\,0.46\,(2)  & 4.59\,$\pm$\,0.45\,(1) & 5.57\,$\pm$\,0.51\,(2) & 4.39\,$\pm$\,0.47\,(2)\\
$_{38}$Sr &                        &                        &                         &                        & 5.26\,$\pm$\,0.49\,(2) & 3.19\,$\pm$\,0.51\,(2)\\
$_{39}$Y  &                        &                        & 4.46\,$\pm$\,0.46\,(2)  & 3.81\,$\pm$\,0.47\,(3) & 4.00\,$\pm$\,0.48\,(7) & 3.19\,$\pm$\,0.41\,(5)\\
$_{56}$Ba &1.69\,$\pm$\,0.55\,(3)  & 1.94\,$\pm$\,0.49\,(2) &  5.08\,$\pm$\,0.65\,(2) & 3.46\,$\pm$\,0.56\,(2) & 3.85\,$\pm$\,0.74\,(1) & 1.68\,$\pm$\,0.50\,(1)\\
\hline
\end{tabular*}
\label{abunresult}
\end{table*}

\begin{figure*}
 \centering
 \begin{minipage}[b]{0.45\textwidth}
  \includegraphics[height=6.5cm, width=1\textwidth]{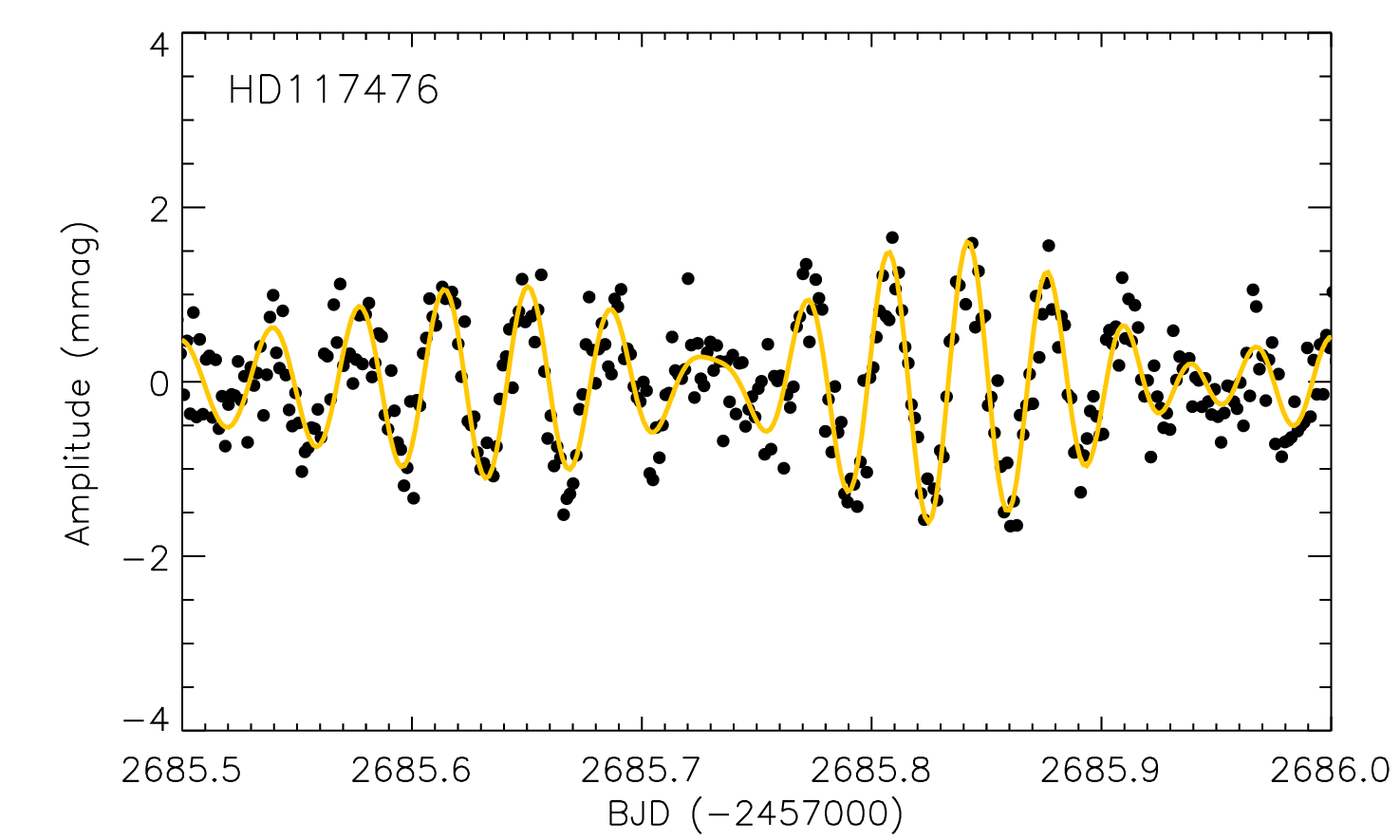}
 \end{minipage}
 \begin{minipage}[b]{0.45\textwidth}
  \includegraphics[height=6.5cm, width=1\textwidth]{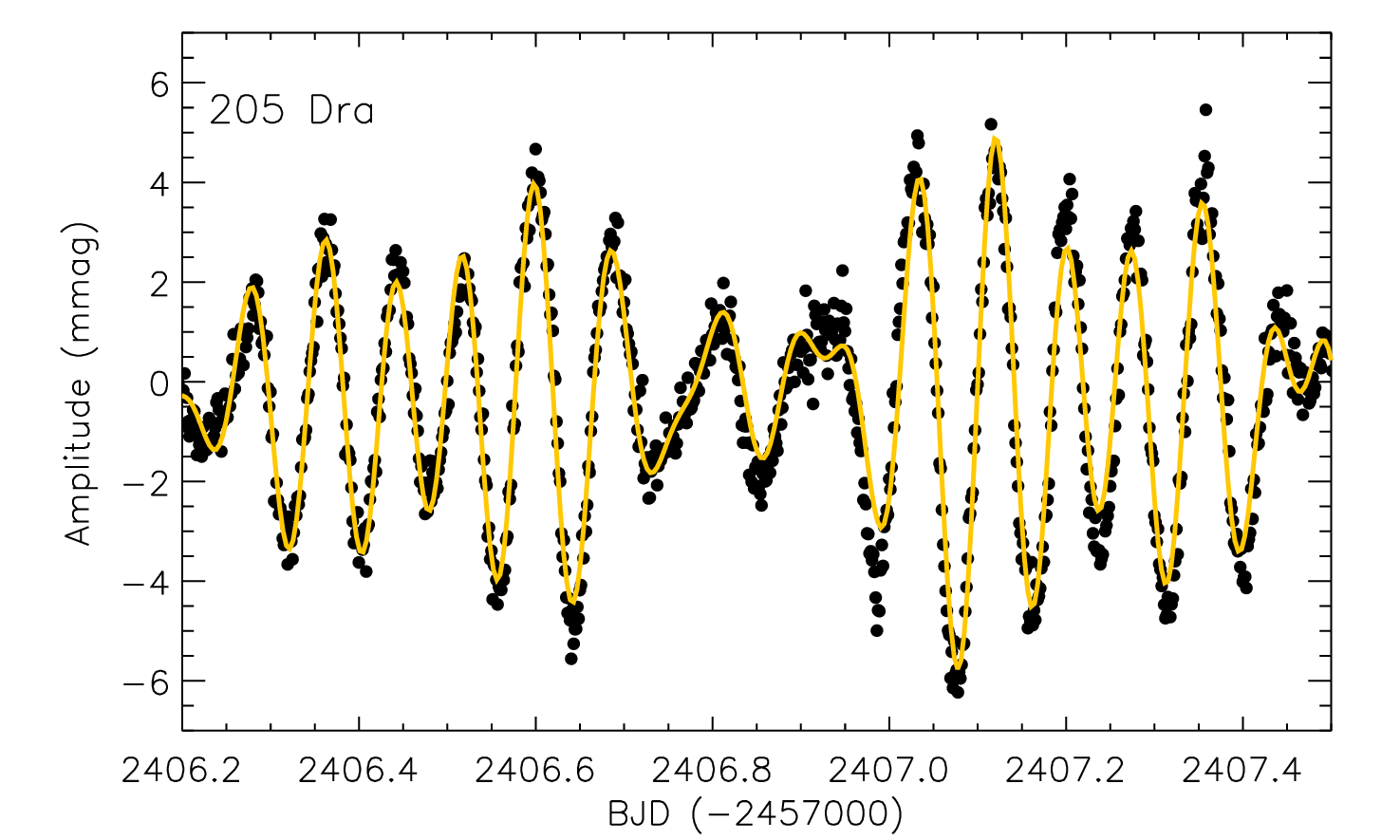}
  \end{minipage}
   \begin{minipage}[b]{0.45\textwidth}
  \includegraphics[height=6.5cm, width=1\textwidth]{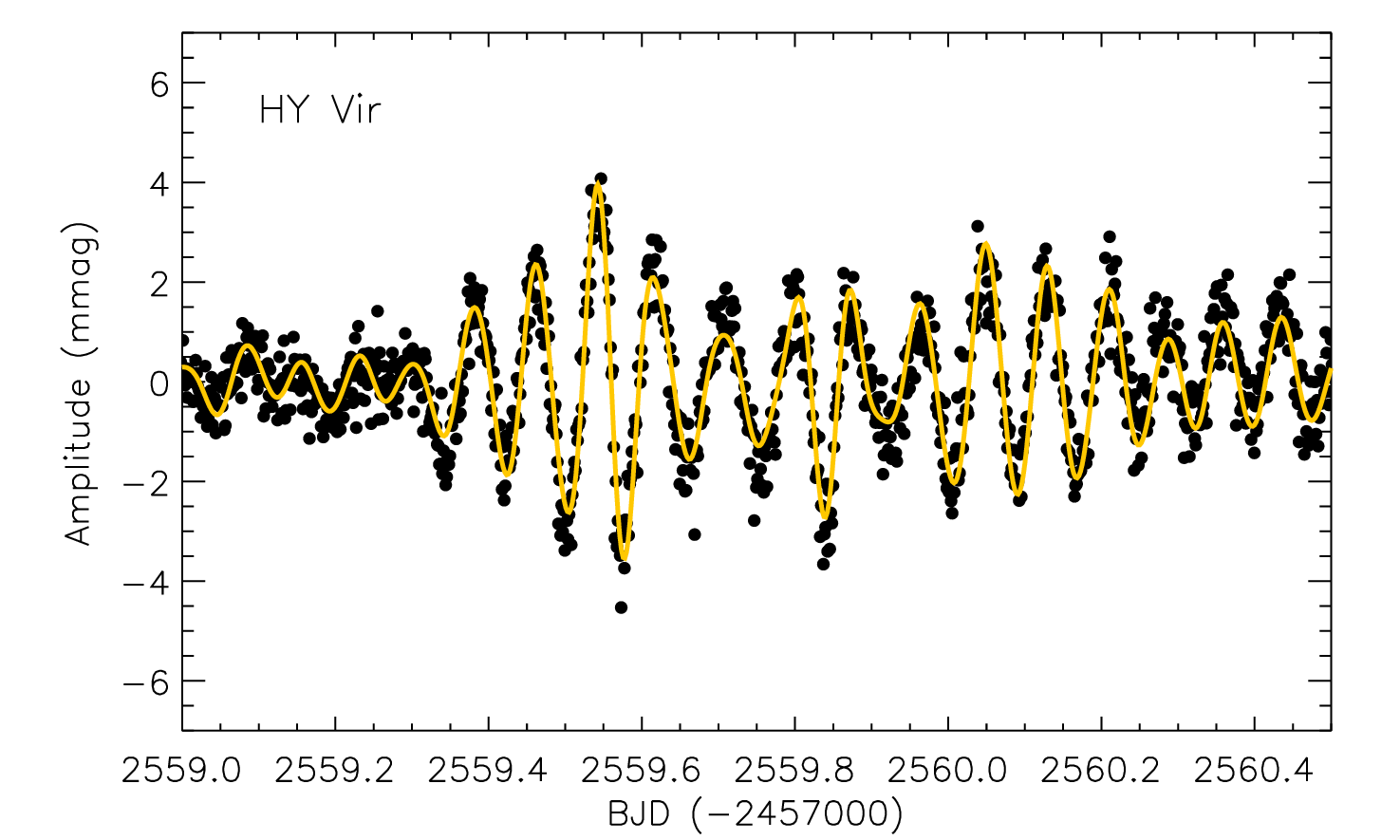}
  \end{minipage}
   \begin{minipage}[b]{0.45\textwidth}
  \includegraphics[height=6.5cm, width=1\textwidth]{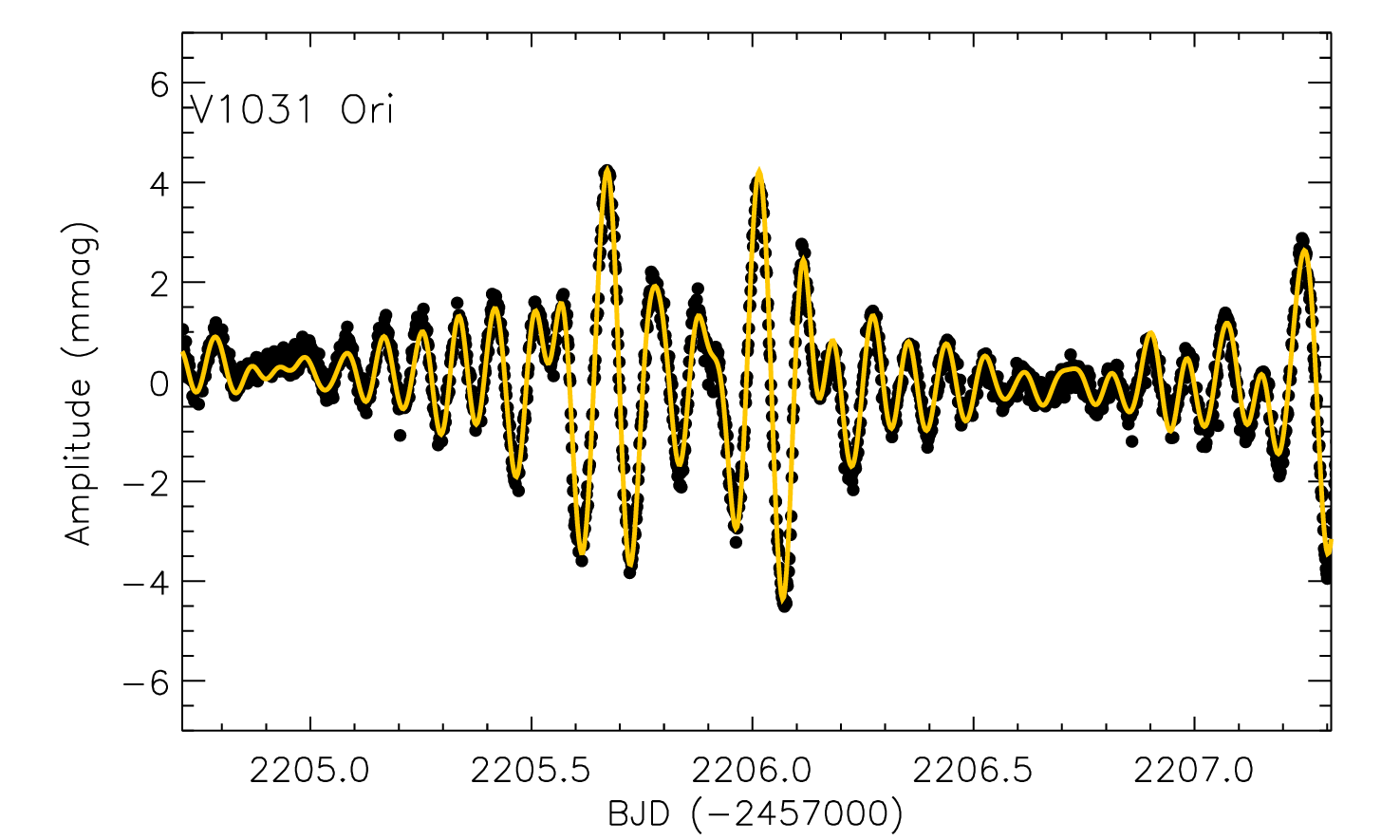}
  \end{minipage}
   \caption{The theoretical pulsation fit to the observations.}  \label{pulsfits}
\end{figure*}

\begin{figure}
  \centering
  \includegraphics[width=1.0\linewidth]{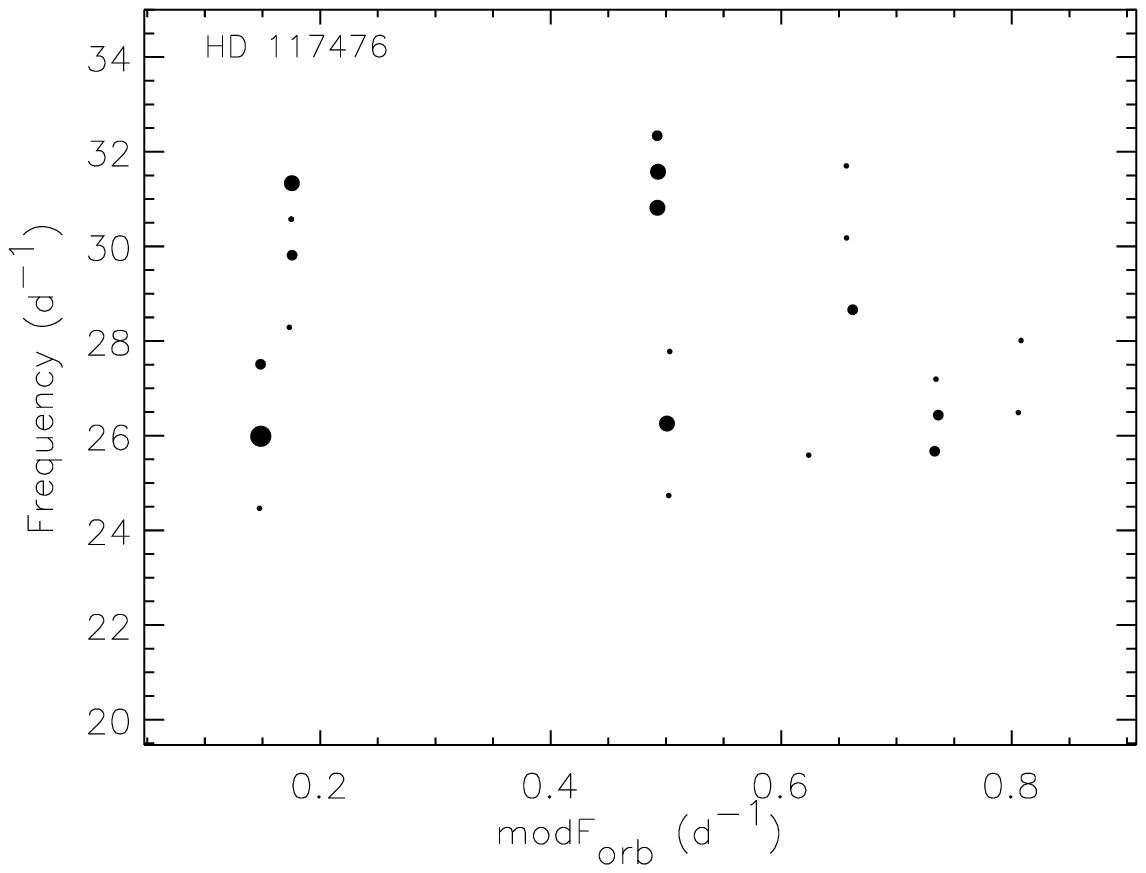}  
\caption{Frequency modulation with the orbital period for HD\,117476. The size of the symbols changes with the amplitude values.}\label{fig2:appen}
\end{figure}



\end{document}